\documentclass[aps,prl,twocolumn,longbibliography]{revtex4-2} 

\usepackage{graphicx}
\usepackage{dcolumn}
\usepackage{bm}
\usepackage{times} 
\usepackage{amsmath,amssymb}
\usepackage{float}
\usepackage{color}
\usepackage{multirow}
\usepackage[normalem]{ulem}
\usepackage{CJK}

\begin{document}

\title{Electron Spin Pumping by Thermal Radiation}

\author{Gaomin Tang} 
\affiliation{Graduate School of China Academy of Engineering Physics, Beijing 100193,
China}

\begin{abstract}
  This Letter demonstrates that thermal radiation can pump angular momentum into the spins
  of itinerant electrons across a vacuum gap, without spatial spin transport or net charge
  flow. In a monolayer transition metal dichalcogenide radiatively coupled to a thermal
  emitter, Rashba spin-orbit coupling enables spin-selective transitions between helical
  conduction bands. Two distinct mechanisms are identified: chiral spin pumping mediated
  by interference between polarization components of thermal radiation from a
  nonreciprocal emitter, and nonchiral spin pumping arising from Zeeman-induced asymmetry
  between opposing spin-flip transitions driven by nonchiral thermal fluctuations.
\end{abstract}

\maketitle

{\it Introduction.}
The generation and control of spin polarization are central goals of
spintronics~\cite{spintronics04,spintronics05}. Thermal driving
offers an alternative to electrical control and underlies spin
caloritronics~\cite{spin-calori-12,spin-calori-26}. Established thermal
spin responses include the spin Seebeck effect mediated by magnons or
phonons~\cite{Seebeck08,Seebeck10,Seebeck10-1,Seebeck10-2,Seebeck13,
Seebeck18,Seebeck_chiral_23,Seebeck_chiral_24,Seebeck_chiral_25,GT26-4},
as well as thermally induced spin polarization in Rashba electron
gases~\cite{Dyrdal18}. Polarization-controlled light provides
another route to spin manipulation through optical orientation~\cite{Meier84}
and the inverse Faraday effect~\cite{Edelstein98,Berritta16,Tanaka24}.
Evanescent microwave fields can pump electron spins without
contact~\cite{YuBauer20}, while electric and magnetic near fields can
jointly drive transverse spin currents~\cite{LiCaiYu26}.
These developments motivate the study of incoherent thermal
radiation as a contactless source of electron spin polarization.

Across subwavelength vacuum gaps, evanescent modes can enhance
radiative heat transfer by orders of magnitude beyond the black-body
limit~\cite{review07,review15,review18,review21}.
Thermal radiation from nonreciprocal media can also carry spin
angular momentum and exert optical torques~\cite{Jacob19,nonrecip_torque21,
Khan19,GT24-2,spin-PBA-26}. For itinerant electrons, spin-orbit
coupling (SOC) allows the electric field of thermal radiation
to drive spin-changing transitions. Net spin pumping requires
an imbalance between transitions that increase and decrease the
electron spin projection.

In this Letter, we identify two mechanisms by which nonequilibrium
thermal radiation can pump electron spins in a monolayer transition
metal dichalcogenide (TMDC) separated from a thermal emitter by a
nanoscale vacuum gap. In both cases, Rashba-induced spin mixing allows thermal
photons to drive spin-selective transitions between helical conduction
bands. When the emitter breaks time-reversal symmetry, chiral
correlations in the radiation favor one direction of spin transfer in
a magnetically unbiased TMDC. Both propagating and evanescent modes
contribute to this chiral pumping rate. When time-reversal symmetry
is broken in the TMDC, Zeeman splitting makes opposing spin-changing
transitions occur at unequal rates even under nonchiral illumination.
Both mechanisms generate a local out-of-plane spin polarization in the TMDC.

\begin{figure}
\centering
\includegraphics[width=\columnwidth]{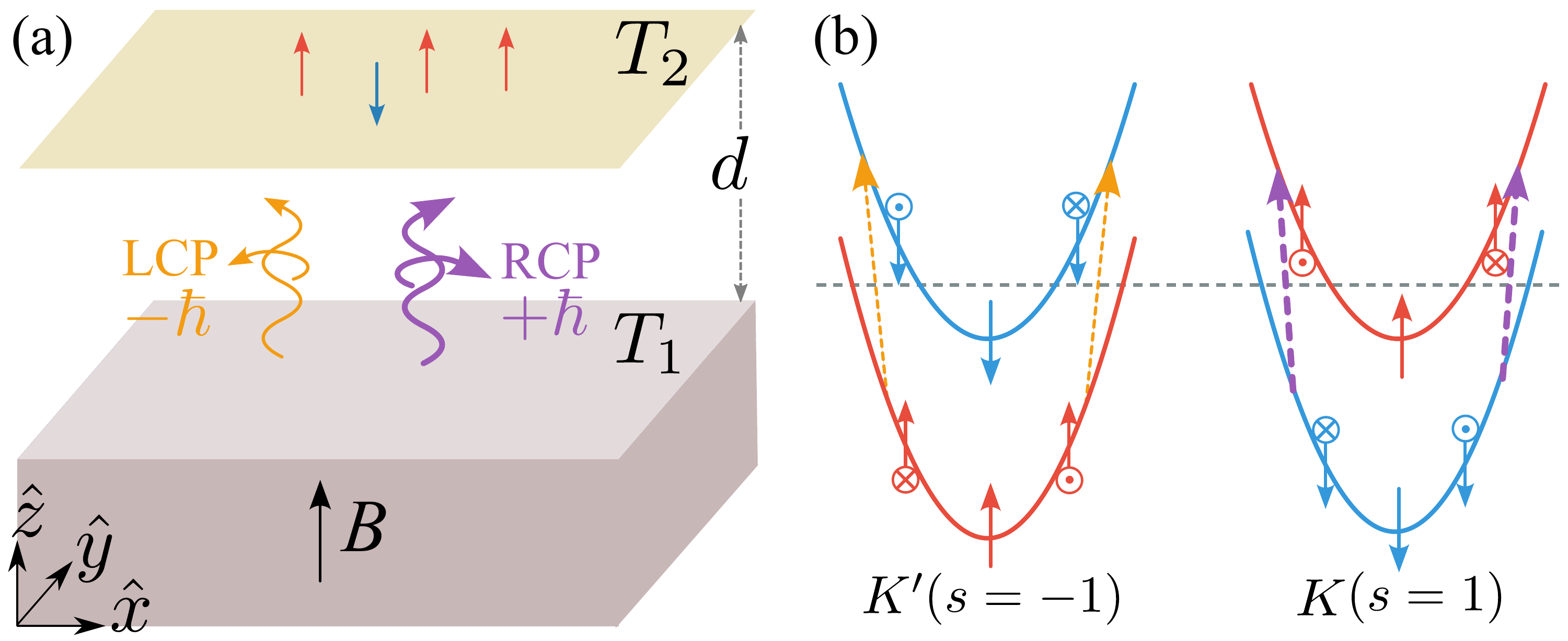} \\
\caption{
  Schematic of electron spin pumping driven by chiral thermal photons.
  (a) A semi-infinite magneto-optic emitter at $T_1$, with optical response
  parameterized by $\bm{B}=B\hat{z}$, couples radiatively across a vacuum
  gap $d$ to a magnetically unbiased monolayer TMDC at $T_2$.
  The emitter produces unequal left- (LCP, orange) and
  right-circularly polarized (RCP, purple) thermal photon fluxes.
  (b) Ising and Rashba SOC produce spin-mixed helical conduction bands,
  with spin orientations illustrated along $\bm{k}=k_x\hat{x}$.
  LCP absorption drives predominantly spin-up (red) to spin-down
  (blue) transitions in valley $K'$, while RCP absorption drives
  the reverse transitions in valley $K$.
  Unequal rates of these opposing transitions generate a 
  local out-of-plane electron spin source.
}
\label{fig1}
\end{figure}

{\it Model and Hamiltonian.}
A semi-infinite emitter at temperature $T_1$ is separated from a heavily electron-doped
TMDC at temperature $T_2$ by a vacuum gap $d$ [Fig.~\ref{fig1}(a)].
Thermal photon energies lie far below the TMDC band gap, restricting
the relevant transitions to the conduction bands.
The Hamiltonian is
$H_{\rm tot}=H_{\rm ph}+H_{\rm e}+H_{\rm int}$, where
$H_{\rm ph}$ describes the photonic environment~\cite{GT26-2,SM}.
The electronic Hamiltonian is 
\begin{equation} \label{He}
  H_{\rm e} = \sum_{\bm{k}} \sum_s C_{\bm{k}}^\dag \left[ \varepsilon_{\bm{k}} \sigma_0 +
  \bm{h}(\bm{k}) \cdot \bm{\sigma} \right] C_{\bm{k}} ,
\end{equation}
with the electron spinor $C_{\bm{k}} = (c_{\bm{k}\uparrow}, c_{\bm{k}\downarrow})^T$.
Here, $\varepsilon_{\bm{k}}= \hbar^2 k^2 / (2m^*)$ is the kinetic energy dispersion
with in-plane wavevector $\bm{k} = (k_x, k_y)$ and effective mass $m^*$. The Pauli
matrices $\sigma_{x,y,z}$ (and the identity matrix $\sigma_0$) act on the electron spin
subspace.  The effective field is
\begin{equation}
  \bm{h}(\bm{k}) = (\lambda_R k_y, -\lambda_R k_x, s\lambda_c + M_z) .
\end{equation}
The intrinsic Ising SOC ($\lambda_c$) is responsible for spin-valley locking, which binds
the out-of-plane spin to the valley index $s=\pm 1$.
The Rashba SOC ($\lambda_R$) produces an in-plane effective field perpendicular to the
electron wavevector.
This arises from out-of-plane structural asymmetry due to gating,
interfaces, or a Janus structure.
Magnetic proximity is represented by the 
out-of-plane exchange term $M_z\sigma_z$, 
which
breaks electronic 
time-reversal symmetry.
Diagonalizing the Hamiltonian yields the eigenenergies
\begin{equation}
  \varepsilon_{n\bm{k},s} = \varepsilon_{\bm{k}} + n \Delta_{\bm{k},s} ,
\end{equation}
where $n=\pm$ labels the helical bands, and the effective field magnitude is given by
$\Delta_{\bm{k},s} =\sqrt{\lambda_R^2 k^2 + (s\lambda_c + M_z)^2}$.
When $M_z=0$, both $\varepsilon_{n\bm{k},s}$ and $\Delta_{\bm{k},s}$ are
valley-independent so that the valley index $s$ explicitly drops out. 
The paramagnetic interaction coupling the electrons to the vector potential
$\bm{A}(\bm{q})$ with in-plane wavevector $\bm{q}$ takes the form
$H_{\text{int}} = -e_0 \sum_{\bm{k}, \bm{q}} C_{\bm{k}+\bm{q}}^\dag \left[
\bm{v}(\bm{k}, \bm{q}) \cdot \bm{A}(\bm{q}) \right] C_{\bm{k}}$,
where $e_0$ is the elementary charge. The symmetrized velocity vertex matrix is given by
$\bm{v}(\bm{k},\bm{q}) = \frac{1}{2} \big[\bm{v}(\bm{k}) + \bm{v}(\bm{k} +
\bm{q})\big]$ (see the End Matter for explicit forms).
It contains the spin-flipping Rashba velocity and a spin-independent diagonal component
governed by the mean kinematic velocity 
\begin{equation}
  \bar{\bm{V}}(\bm{k}, \bm{q}) = \hbar ( \bm{k} + \bm{q}/2 ) /  m^* .
\end{equation}

{\it Spin pumping by chiral thermal photons.}
We first consider an emitter magnetized out of plane and a time-reversal-symmetric
receiver ($M_z=0$).
In an in-plane isotropic emitter, out-of-plane magnetization
preserves rotational symmetry while producing unequal circular
photon correlations. These correlations drive opposing
spin-changing transitions at different rates, generating an
out-of-plane spin source [Fig.~\ref{fig1}].

The required chiral photon imbalance can also be generated from the intrinsic magneto-optic
response of magnetic Weyl semimetals~\cite{WSM_SPP16, Kotov18, WSM_AHE_20nc, GT_WSM} or 2D
ferromagnetic insulators~\cite{2D_magnets_26}. Unlike traditional
three-dimensional ferromagnets with low magnon frequencies, 2D van der Waals ferromagnets
(e.g., CrI$_3$) host anisotropy-driven terahertz magnon polaritons capable of bridging
interband spin-flip transitions~\cite{2D_magnets_26}. Furthermore, using
such magnetic insulators provides an additional capability: beyond static temperature
differences, one can use dynamic spin pumping by periodically modulating the emitter's
magnetization~\cite{SP02, SP25}, a mechanism closely related to Floquet
thermal radiation~\cite{Li19, Li21, Biehs22, GT24, YuFan24, GT25}.

Evaluating the leading-order electron-photon collision integral within the
nonequilibrium Green's function formalism~\cite{Lifshitz_book, Haug_Jauho,
JSW23, JSW25, GT26-2}, we obtain the chiral spin pumping rate including
both evanescent and propagating modes~\cite{SM}:
\begin{equation}
  I_z^{\rm ch} = \frac{\hbar}{2}\int_0^\infty d\omega
  \int\frac{d^2\bm{q}}{(2\pi)^2}\frac{\mu_0}{k_0}
  \Phi_{\rm ch}(\bm{q},\omega)N_d(\omega)\chi_{\rm ch}(\bm{q},\omega),
\end{equation}
where $k_0=\omega/c$ is the vacuum wavenumber, $c$ is the speed of light
in vacuum, and $\mu_0$ is the vacuum permeability.
Here, $I_z^{\rm ch}$ denotes the local rate per unit area at which
the electrons acquire out-of-plane spin angular momentum through
chiral pumping. It is positive when the transitions produce a net
increase in the electron spin projection along $z$.
The rate is determined by the electronic response kernel $\chi_{\rm ch}$,
the chiral photon flux factor $\Phi_{\rm ch}$, and the thermal occupation
difference $N_d$. With the radiation incident from infinity at the receiver 
temperature $T_2$, the occupation difference is
\begin{equation} \label{Nd}
  N_d(\omega)=[\exp(\hbar\omega/k_BT_1)-1]^{-1}
  -[\exp(\hbar\omega/k_BT_2)-1]^{-1}.
\end{equation}
The photon flux factor separates into
$\Phi_{\rm ch}^{\rm EV}$ for evanescent modes ($q>k_0$) and
$\Phi_{\rm ch}^{\rm PR}$ for propagating modes ($q<k_0$).
To express these contributions, we introduce the reflection coefficients
$r_1$ for the emitter and $r_2$ for the TMDC, together with the TMDC
transmission coefficients $t_2$. The emitter coefficients are calculated
following Ref.~\cite{GT_WSM}, whereas the TMDC coefficients are derived
in the End Matter.
Multiple reflections across the vacuum gap are characterized by
\begin{equation}
  \Delta_{s,p}=1-r_1^{s,p}r_2^{s,p}e^{2i\gamma_0d},
\end{equation}
where $\gamma_0=\sqrt{k_0^2-q^2}$ is chosen with nonnegative real and
imaginary parts. The factor $e^{2i\gamma_0d}$ therefore describes exponential
decay in the evanescent sector and phase accumulation in the propagating
sector. For the isotropic, magnetically unbiased TMDC considered here,
the sheet response is diagonal in the $s,p$ basis.
Retaining terms to first order in the emitter's cross-polarization
reflection coefficients gives~\cite{SM}
\begin{align}
  \Phi_{\rm ch}^{\rm EV}(& \bm{q}, \omega) = {\rm Im} \Big\{ t_2^p (t_2^s)^* \Big[ 
  (r_1^{sp})^* \big[1 - r_1^p (r_2^p)^* e^{-2|\gamma_0|d}\big]\Delta_s   \notag \\
  & - r_1^{ps} \big[1 - (r_1^s)^* r_2^s e^{-2|\gamma_0|d}\big] \Delta_p^* \Big] \Big\} 
  \frac{e^{-2|\gamma_0| d}}{|\Delta_p|^2 |\Delta_s|^2} , \label{Phi_ch_EV}
\end{align}
for $q>k_0$, and
\begin{align} 
  \Phi_{\rm ch}^{\rm PR}
  = & {\rm Im}\Big\{t_2^p(t_2^s)^* \Big[(r_1^{sp})^*
  \big(r_1^p-(r_2^p)^*e^{-2i\gamma_0d}\big) \Delta_s  \notag \\
  & +r_1^{ps} \big((r_1^s)^*-r_2^se^{2i\gamma_0d}\big) \Delta_p^*
  \Big]\Big\} \frac{1}{|\Delta_p|^2|\Delta_s|^2} , \label{Phi_ch_PR}
\end{align}
for $q<k_0$. Both expressions describe phase-sensitive correlations
between the in-plane $s$- and $p$-polarized electric fields:
$r_1^{sp}$ and $r_1^{ps}$ describe polarization conversion at the emitter,
while $t_2^p(t_2^s)^*$ accounts for the relative transmission amplitudes
and phases at the sheet.
With the adopted polarization convention, $\Phi_{\rm ch}$ encodes the
imbalance between the RCP and LCP components of the in-plane electric
field~\cite{SM}.

For the isotropic electronic model considered here, the chiral response kernel is
\begin{equation}
  \chi_{\rm ch}(\bm{q},\omega)
  = \frac{2e_0^2}{\pi}
  \int \frac{d^2\bm{k}}{(2\pi)^2} \sum_{n,m}
  \mathrm{Re}\big[\Lambda^{\rm ch}_{nm}\big]\,
  \mathcal{L}_{mn}(\bm{k},\bm{q},\omega),
\end{equation}
where $n,m=\pm$ label the helicity bands, and the factor of $2$ accounts for the two
valleys. For $\eta\ll k_BT_2$, the absorptive transition kernel is
\begin{equation} \label{Lmn}
  \mathcal{L}_{mn}(\bm{k},\bm{q},\omega)
  \simeq -\mathrm{Im}\left[
  \frac{f(\varepsilon_{n\bm{k}}+i\eta)
  -f(\varepsilon_{m,\bm{k}+\bm{q}}-i\eta)}
  {\varepsilon_{nm}+\hbar\omega+2i\eta}
  \right],
\end{equation}
where $\eta$ is the electronic broadening,
$\varepsilon_{nm}=\varepsilon_{n\bm{k}}
-\varepsilon_{m,\bm{k}+\bm{q}}$, and
$f(z)=\{\exp[(z-\mu)/(k_BT_2)]+1\}^{-1}$ is the analytically
continued Fermi-Dirac distribution at chemical potential $\mu$.
The derivation is detailed in the Supplemental Material~\cite{SM}.
Finally, the chiral transition vertex evaluates to
\begin{align}
  \mathrm{Re}[\Lambda^{\rm ch}_{nm}(\bm{k},\bm{q})] = & \frac{\lambda_R^2}{2\hbar^2}
  \bar{\bm{V}}\cdot \left[ \frac{n\hbar\bm{k}}{\Delta_{\bm{k}}} +
  \frac{m\hbar(\bm{k}+\bm{q})}{\Delta_{\bm{k}+\bm{q}}}  \right] \notag \\
  &+ \frac{\lambda_R^2}{\hbar^2} \left( 1 - \frac{n m \lambda_c^2}{\Delta_{\bm{k}}
  \Delta_{\bm{k}+\bm{q}}} \right) .
\end{align}
This explicit $\lambda_R^2$ dependence confirms that Rashba SOC is required for the
photon-induced spin pumping.

In the numerical calculation, the thermal emitter is modeled as $n$-doped InSb with a
gyrotropic Drude dielectric tensor parameterized by an out-of-plane magnetic field $B$, as
detailed in the End Matter. The gyrotropic response originates from the cyclotron motion
of free carriers, which generates off-diagonal components in the dielectric tensor.
Alternatively, a remanently magnetized magneto-optical medium could provide a
nonreciprocal optical response without external magnetic bias. To isolate the chiral
spin-pumping mechanism, the TMDC receiver is assumed to be magnetically unbiased.

The emitter and receiver are separated by a vacuum gap of $d=20\,$nm and maintained at
temperatures $T_1=330\,$K and $T_2=300\,$K, respectively. For the electron-doped
TMDC receiver, we adopt a chemical potential $\mu = 100\,$meV, an effective mass $m^* =
0.4m_0$ with $m_0$ the free electron mass, and an Ising SOC strength $\lambda_c=12\,$meV.
The phenomenological electronic broadening is set to $\eta=4\,$meV. Taking $\tau_p
=\hbar/(2\eta)$ gives a Drude mobility of $\mu_{\rm mob}=e_0\tau_p/m^* \approx
360\,{\rm cm}^2/({\rm V\,s})$. 
Using in-plane rotational symmetry, we take $\bm{q}=q\hat{x}$ and integrate over the
electron wavevector $\bm{k}$ in polar coordinates.
The photon wavevector integration extends to $q_{\max}=1/d$, with convergence verified.
The characteristic Fermi wavevector is $k_F = \sqrt{2m^* \mu}/\hbar \approx
10^9\,$m$^{-1}$, corresponding to a transition gap of $2\sqrt{\lambda_R^2
k_F^2+\lambda_c^2}\approx25\,$meV for $\lambda_R=40\,$meV$\cdot$\AA.

\begin{figure}
\centering
\includegraphics[width=\columnwidth]{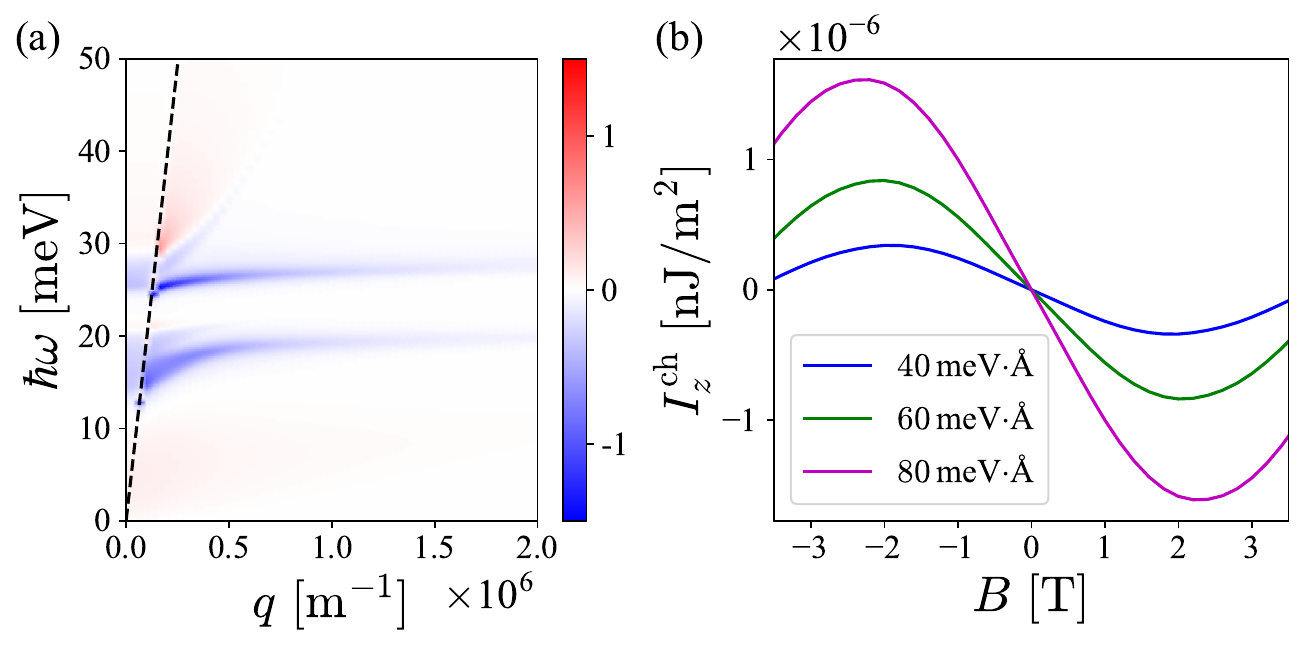} \\
\caption{(a) Chiral photon flux factor $\Phi_{\rm ch}(\bm{q}, \omega)$ versus
  in-plane wavevector magnitude $q$ and photon energy $\hbar\omega$
  for an emitter magnetic field of $B=2\,$T and a Rashba SOC strength
  of $\lambda_R=40\,$meV$\cdot$\AA.
  The color scale quantifies the spectral imbalance between RCP (red) and LCP (blue) 
  thermal photons. The dashed line represents the light line.
  (b) Spin pumping rate $I_z^{\rm ch}$ as a function of $B$
  for varying $\lambda_R$. }
\label{fig2}
\end{figure}

Figure~\ref{fig2}(a) maps the chiral photon flux factor $\Phi_{\rm ch}(\bm{q}, \omega)$
at $B = 2\,$T and $\lambda_R=40\,$meV$\cdot$\AA. 
Despite the broad emission spectrum, the electronic response kernel preferentially
weights photons near the spin-flip transition energies, with a finite linewidth due to
electronic scattering. The spin pumping rate is thus determined by integrating the chiral
photon flux over both propagating and evanescent modes with the thermal and
electronic response weights. Here, the counterclockwise-rotating electric field of the
RCP mode couples with the Lorentz-driven electron cyclotron orbits, pushing the upper RCP
dispersion branch into the radiative bulk continuum. While the LCP mode governs the local
flux near the characteristic spin-flip energy of $25\,$meV, the RCP mode takes over
at higher energies.
After weighting by the thermal occupation difference and the electronic response,
the LCP contribution outweighs the RCP contribution, yielding a negative spin pumping
rate.

Figure~\ref{fig2}(b) shows the spin pumping rate versus the magnetic field $B$. 
The rate is odd in $B$ because field reversal reverses the cyclotron motion and
interchanges the LCP and RCP emission profiles.
In addition, the pumping rate displays a nonmonotonic dependence on $B$. Initially,
its magnitude increases with the magnetic field due to the increased chiral
asymmetry of the photon flux. However, the magnitude peaks near $B=2\,$T and then
decreases, consistent with a field-induced shift of the emission resonances relative to
the fixed spin-flip transition energies of the TMDC.
Although enhanced spin mixing drives a monotonic increase in the magnitude of the
pumping rate across the plotted $\lambda_R$ range, this magnitude is expected
to decay at large $\lambda_R$ because strong SOC both detunes the spin-flip transition
energy from the thermal emission and forces the electron spins into the plane.

\begin{figure}
\centering
\includegraphics[width=\columnwidth]{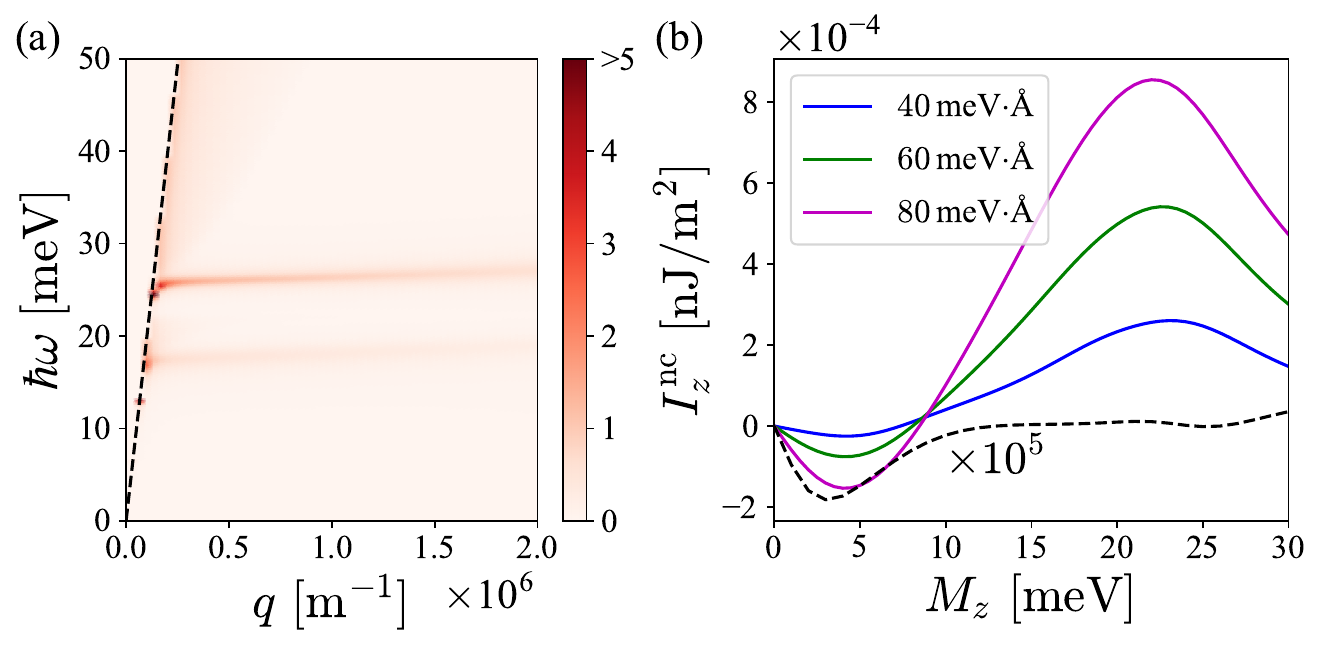} \\
\caption{(a) Nonchiral photon flux factor $\Phi_{\rm nc}(\bm{q}, \omega)$ versus
  in-plane wavevector $q$ and photon energy $\hbar\omega$ for an exchange
  parameter of
  $M_z = 0.5\,$meV and Rashba SOC strength $\lambda_R=40\,$meV$\cdot$\AA.  
  (b) Nonchiral spin pumping rate $I_z^{\rm nc}$ (solid lines) as a function of $M_z$ for
  various $\lambda_R$. 
  The dashed line, scaled by a factor of $10^5$, represents the chiral pumping rate
  at $\lambda_R=80\,$meV$\cdot$\AA.  
}
\label{fig3}
\end{figure}

{\it Spin pumping by nonchiral thermal photons.}
We now turn to the regime where a finite out-of-plane exchange parameter ($M_z \neq 0$) breaks time-reversal
symmetry within the 2D electron gas. Here, the thermal emitter generates a nonchiral
photon flux.
The finite exchange parameter induces a Zeeman splitting between the spin-up and spin-down
electrons. By breaking the balance between the opposing spin-flip transitions ($\uparrow
\to \downarrow$ versus $\downarrow \to \uparrow$), this energy separation causes the
nonchiral thermal fluctuations to drive the transitions at unequal rates, rectifying the
photon flux into a net out-of-plane spin pumping rate.
In a proximity-coupled device, the magnetic layer would directly contact the back side of
the TMDC, leaving its emitter-facing surface exposed to the vacuum gap. Quantifying the
magnetic layer's effect on the effective Fresnel coefficients requires modeling the
optical response of the full layer stack.
The chiral photon flux remains finite due to polarization conversions within the 2D
electron gas (see the End Matter). However, its contribution is negligible compared with
the nonchiral contribution as shown in Fig.~\ref{fig3}(b). 

For the suspended TMDC considered here,
the spin pumping rate driven by nonchiral near-field thermal photons is obtained
as~\cite{SM}
\begin{equation} \label{Iz_nc}
  I_z^{\rm nc} = \frac{\hbar}{2} \int_0^\infty d\omega \int_{\rm EV}
  \frac{d^2\bm{q}}{(2\pi)^2} \frac{|\gamma_0|}{\epsilon_0 \omega^2} 
  \Phi_{\rm nc}(\bm{q}, \omega) N_d(\omega) \chi_{\rm nc}(\bm{q}, \omega) ,
\end{equation}
where $\epsilon_0$ is the vacuum permittivity. The photonic environment is captured by the
nonchiral photon flux factor, which describes the $p$-polarized evanescent wave reaching
the 2D metal~\cite{GT26-2}:
\begin{equation}
  \Phi_{\rm nc}(\bm{q}, \omega) = |t_2^p|^2 \text{Im}(r_1^p) e^{-2|\gamma_0|d}
  \big/ |\Delta_p|^2 .
\end{equation}
We retain the dominant $p$-polarized evanescent contribution, since nonchiral pumping
benefits from near-field enhancement without requiring phase-sensitive interference
between the $s$- and $p$-polarized electric fields.
The electronic response is described by the nonchiral response kernel,
\begin{equation}
  \chi_{\rm nc}(\bm{q},\omega)
  = \frac{e_0^2}{\pi}
  \int \frac{d^2\bm{k}}{(2\pi)^2} \sum_{n,m,s}
  \mathrm{Im}\big[\Lambda^{\rm nc}_{nm,s}\big]\,
  \mathcal{L}_{mn,s}(\bm{k},\bm{q},\omega),
\end{equation}
where $s=\pm1$ labels the valleys and $\mathcal{L}_{mn,s}$ is the
valley-resolved absorptive transition kernel defined in Eq.~\eqref{Lmn}.
The transition energies and occupation factors are valley dependent due to finite $M_z$.
The nonchiral transition vertex is
\begin{align}
  & \mathrm{Im}[\Lambda^{\rm nc}_{nm, s}(\bm{k},\bm{q})] = - \frac{\lambda_R^2}{\hbar^2}
  (s\lambda_c + M_z) \times \notag \\
  & \left( \frac{m n \hbar}{2 \Delta_{\bm{k},s} \Delta_{\bm{k}+\bm{q},s}} \bar{\bm{V}}
  \cdot \bm{q} + \frac{n}{\Delta_{\bm{k},s}} - \frac{m}{\Delta_{\bm{k}+\bm{q},s}} \right).
\end{align}
At $M_z=0$, the vertex is valley odd while the energies are
valley degenerate, so the valley contributions cancel.
The resulting pumping rate is odd in $M_z$.

Figure~\ref{fig3} shows the nonchiral photon flux factor and the
spin pumping rate. The emitter is modeled as $n$-doped
InSb with the same material parameters as in Fig.~\ref{fig2},
but with the applied magnetic field set to zero.
The two peaks near $17$ and $26\,$meV in Fig.~\ref{fig3}(a)
originate from the surface polariton modes of InSb.

\begin{figure}
\centering
\includegraphics[width=\columnwidth]{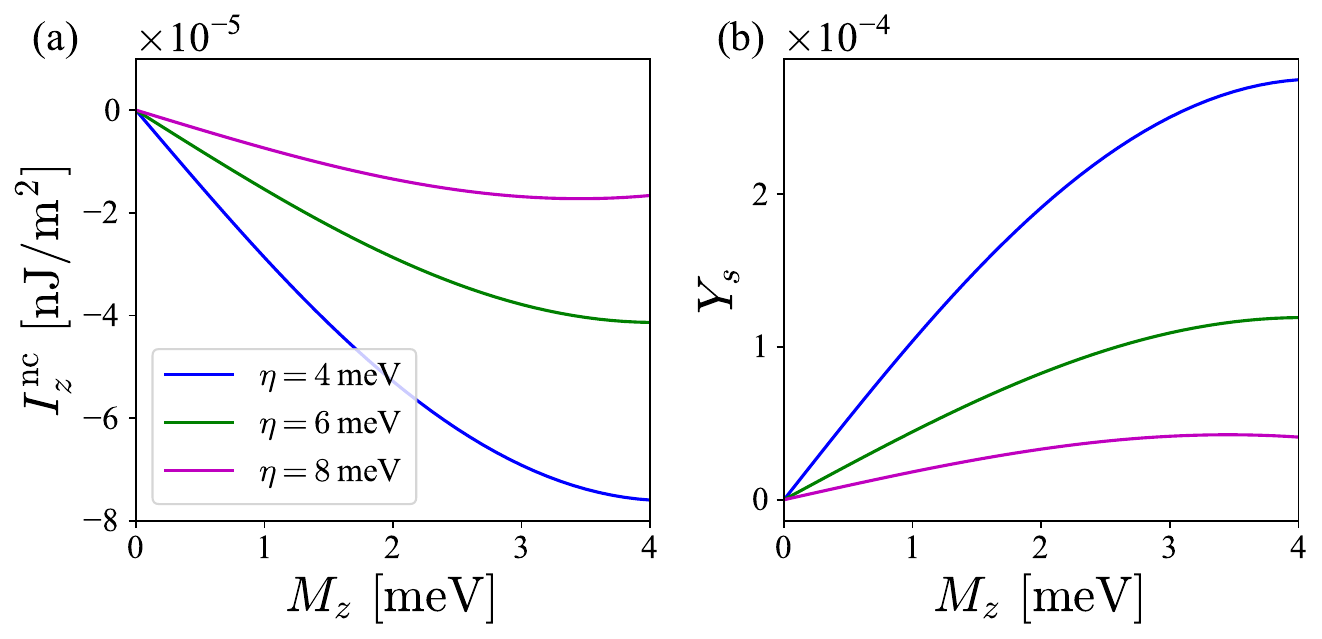} \\
\caption{
  (a) Nonchiral spin pumping rate $I_z^{\rm nc}$ and
  (b) net spin yield $Y_s = |I_z^{\rm nc}| / (\hbar\dot{N}_{\rm ph})$
  versus $M_z$ for different electronic broadenings $\eta$.
  The broadenings $\eta=4$, $6$, and $8\,$meV correspond to estimated Drude
  mobilities of $360$, $240$, and $180\,\mathrm{cm^2/(V\,s)}$, respectively.
  The parameters are the same as in Fig.~\ref{fig3}, with
  $\lambda_R=60\,$meV$\cdot$\AA.
}
\label{fig4}
\end{figure}

Figure~\ref{fig3}(b) shows the nonchiral spin pumping rate versus
the out-of-plane exchange parameter $M_z$. Its nonmonotonic
dependence reflects changes in both the spectral overlap with
the near-field photon flux and the balance between opposing
spin-changing transitions.
For small positive $M_z$, the negative contribution from the
$K'$ valley outweighs the positive contribution from the
$K$ valley, yielding a negative net pumping rate. 
For $\lambda_R=40\,$meV$\cdot$\AA, the negative
extremum occurs near $M_z=4.5\,$meV, where the characteristic
spin-flip energy
$2\sqrt{\lambda_R^2 k_F^2+(M_z-\lambda_c)^2}\simeq 17\,$meV
lies near the lower photon-flux peak.
Further increasing $M_z$ changes the relative valley contributions,
which cancel at the sign reversal. The rate subsequently becomes
positive and reaches a maximum as the spin-flip transition energies
in the $K'$ valley overlap with the upper photon-flux peak.
Beyond this maximum, increasing $M_z$ detunes these transitions
from the peak, reducing the spectral overlap and suppressing
the pumping rate.
Figure~\ref{fig4}(a) in the End Matter shows that the
nonchiral pumping rate decreases in magnitude as $\eta$ increases.

The broad $M_z$ scan illustrates the sign reversal and resonant
enhancement of the pumping rate. However, the exchange strengths
near the maximum remain to be established experimentally for TMDC
proximity structures at the temperature considered here.
We therefore focus on the weak-exchange regime.
At $M_z=0.5\,$meV and $\lambda_R=80\,$meV$\cdot$\AA, for example,
the rate is approximately $-3\times10^{-5}\,\mathrm{nJ/m^2}$,
about $3.5\%$ of the peak magnitude at the same Rashba coupling.

The enhanced $p$-polarized evanescent field helps explain why the nonchiral pumping rates
are generally larger in magnitude than the chiral rates in Fig.~\ref{fig2}.
Chiral pumping requires out-of-phase correlations between $s$- and $p$-polarized
components, whose enhancement is limited by the weak $s$ component in the deeply
evanescent regime.

{\it Discussion and conclusion.}
In Ref.~\cite{GT26-2}, nonreciprocal surface plasmons drive a charge current by
  transferring in-plane momentum to electrons. In the rotationally symmetric
  configurations here, the net in-plane momentum transfer vanishes, yet spin-selective
transitions between Rashba-mixed bands can transfer out-of-plane spin angular momentum to
the electrons. 

In the relaxation-time approximation, spin pumping balances spin
relaxation in steady state, yielding a spin accumulation
$\delta S_z \simeq I_z\tau_s$ and a dimensionless polarization
$P_z \simeq 2I_z\tau_s/(\hbar n_{2\rm D})$, where $\tau_s$ is the effective
spin relaxation time and $n_{2\rm D}$ is the carrier density~\cite{SM}.
We consider the nonchiral regime at $M_z=0.5\,$meV and $\lambda_R=80\,$meV$\cdot$\AA,
where the $K'$ valley dominates. 
To estimate the relaxation timescale, we consider
intravalley D'yakonov-Perel' (DP) relaxation: momentum scattering randomizes the
in-plane Rashba field $\Omega_R = 2\lambda_R k_F/\hbar$, while the
out-of-plane field $\Omega_I = 2|s\lambda_c+M_z|/\hbar$ suppresses spin
relaxation~\cite{DP72, DP13, DP17}. For the dominant valley,
$|s\lambda_c+M_z|=11.5\,$meV. For $\eta=4\,$meV, $\tau_p \simeq 0.082\,$ps, and the DP
expression $\tau_s^{\rm DP} \simeq (1+\Omega_I^2\tau_p^2)/(\Omega_R^2\tau_p)$ gives
$\tau_s^{\rm DP}\simeq 0.2\,$ps. 
Since $\Omega_R\tau_p\simeq 2$
lies outside the weak-precession regime assumed in the derivation,
we use this lifetime as an order-of-magnitude estimate.
With $I_z^{\rm nc}\simeq-3\times10^{-14}\,\mathrm{J/m^2}$ from
Fig.~\ref{fig3}(b) and $n_{2\rm D}\simeq3.3\times10^{13}\,\mathrm{cm}^{-2}$,
this gives $P_z^{\rm nc}\sim-3\times10^{-10}$.
Its negative sign indicates
that the excess electron spin points along $-\hat{z}$. For comparison,
assuming a longer effective lifetime of $\tau_s=1\,$ns at the same pumping
rate gives $P_z^{\rm nc} \sim -2\times10^{-6}$.
The End Matter also quantifies the net spin transferred per net absorbed
photon through the spin yield $Y_s$ [Fig.~\ref{fig4}(b)].

To place the estimated polarization in context, we consider the
reversible limit for a spin-flip transition in the dominant $K'$ valley.
At the same $M_z$ and $\lambda_R$, the characteristic transition energy is
$\hbar\omega \simeq 2\sqrt{\lambda_R^2 k_F^2+(M_z-\lambda_c)^2}
\simeq 28\,$meV. For $T_1=330\,$K and $T_2=300\,$K, the maximum spin
chemical potential splitting for this transition is
$\Delta\mu_s^{\max}=\hbar\omega(1-T_2/T_1)\simeq 2.6\,$meV.
In the constant-density-of-states approximation, the corresponding
polarization magnitude is $P_z^{\max}=\Delta\mu_s^{\max}/(2\mu)
\simeq 1.3\%$. The estimated $|P_z^{\rm nc}|$ lies well below this
scale because rapid spin relaxation limits the polarization that
can accumulate under weak radiative pumping.

Polar Kerr rotation could probe the pump-induced out-of-plane spin
polarization~\cite{Yang15,Dey17}, while helicity-resolved reflectance
could distinguish changes in the optical response of the two
valleys~\cite{Back17}. In the weak-exchange regime considered here,
detection would require resolving a polarization change of
$|P_z^{\rm nc}|\sim3\times10^{-10}$, corresponding to an effective spin
density imbalance of $2|\delta S_z|/\hbar\sim10^4\,\mathrm{cm}^{-2}$.
Assessing whether this signal is detectable requires calibrating the
Kerr rotation or reflectance contrast of the doped TMDC on its
substrate. 

In conclusion, this work establishes two mechanisms for contactless
spin pumping by thermal radiation. Rashba SOC enables
spin-selective electronic transitions, while either chiral photon
correlations or Zeeman-induced asymmetry between transitions driven
by a nonchiral photon field generates a local out-of-plane spin source.
The resulting steady-state spin accumulation depends on relaxation
and is small for the parameters considered here. Engineering the
emission spectrum to improve its overlap with spin-changing
transitions offers a route to optimize the spin yield, while
suppressing spin relaxation is essential for enhancing the
accumulated polarization.
By connecting fluctuational electrodynamics with microscopic spin-dependent electronic
transitions, this work identifies a contactless route for transferring angular momentum
from thermal radiation to the spins of itinerant electrons.

{\it Acknowledgments.}
G.T. thanks J. Liu and D. He for helpful discussions and the
anonymous referees for constructive comments that improved the manuscript. 
This work is supported by Science Challenge Project (Grant No. TZ2025017) and National
Natural Science Foundation of China (Grants Nos. 12374048 and 12688201).

\bibliography{bib_heat_radiation}

\begin{thebibliography}{61}%
\makeatletter
\providecommand \@ifxundefined [1]{%
 \@ifx{#1\undefined}
}%
\providecommand \@ifnum [1]{%
 \ifnum #1\expandafter \@firstoftwo
 \else \expandafter \@secondoftwo
 \fi
}%
\providecommand \@ifx [1]{%
 \ifx #1\expandafter \@firstoftwo
 \else \expandafter \@secondoftwo
 \fi
}%
\providecommand \natexlab [1]{#1}%
\providecommand \enquote  [1]{``#1''}%
\providecommand \bibnamefont  [1]{#1}%
\providecommand \bibfnamefont [1]{#1}%
\providecommand \citenamefont [1]{#1}%
\providecommand \href@noop [0]{\@secondoftwo}%
\providecommand \href [0]{\begingroup \@sanitize@url \@href}%
\providecommand \@href[1]{\@@startlink{#1}\@@href}%
\providecommand \@@href[1]{\endgroup#1\@@endlink}%
\providecommand \@sanitize@url [0]{\catcode `\\12\catcode `\$12\catcode
  `\&12\catcode `\#12\catcode `\^12\catcode `\_12\catcode `\%12\relax}%
\providecommand \@@startlink[1]{}%
\providecommand \@@endlink[0]{}%
\providecommand \url  [0]{\begingroup\@sanitize@url \@url }%
\providecommand \@url [1]{\endgroup\@href {#1}{\urlprefix }}%
\providecommand \urlprefix  [0]{URL }%
\providecommand \Eprint [0]{\href }%
\providecommand \doibase [0]{https://doi.org/}%
\providecommand \selectlanguage [0]{\@gobble}%
\providecommand \bibinfo  [0]{\@secondoftwo}%
\providecommand \bibfield  [0]{\@secondoftwo}%
\providecommand \translation [1]{[#1]}%
\providecommand \BibitemOpen [0]{}%
\providecommand \bibitemStop [0]{}%
\providecommand \bibitemNoStop [0]{.\EOS\space}%
\providecommand \EOS [0]{\spacefactor3000\relax}%
\providecommand \BibitemShut  [1]{\csname bibitem#1\endcsname}%
\let\auto@bib@innerbib\@empty
\bibitem [{\citenamefont {{\v{Z}}uti{\'c}}\ \emph {et~al.}(2004)\citenamefont
  {{\v{Z}}uti{\'c}}, \citenamefont {Fabian},\ and\ \citenamefont
  {Sarma}}]{spintronics04}%
  \BibitemOpen
  \bibfield  {author} {\bibinfo {author} {\bibfnamefont {I.}~\bibnamefont
  {{\v{Z}}uti{\'c}}}, \bibinfo {author} {\bibfnamefont {J.}~\bibnamefont
  {Fabian}},\ and\ \bibinfo {author} {\bibfnamefont {S.~D.}\ \bibnamefont
  {Sarma}},\ }\bibfield  {title} {\bibinfo {title} {Spintronics: Fundamentals
  and applications},\ }\href@noop {} {\bibfield  {journal} {\bibinfo  {journal}
  {Rev. Mod. Phys.}\ }\textbf {\bibinfo {volume} {76}},\ \bibinfo {pages} {323}
  (\bibinfo {year} {2004})}\BibitemShut {NoStop}%
\bibitem [{\citenamefont {Tserkovnyak}\ \emph {et~al.}(2005)\citenamefont
  {Tserkovnyak}, \citenamefont {Brataas}, \citenamefont {Bauer},\ and\
  \citenamefont {Halperin}}]{spintronics05}%
  \BibitemOpen
  \bibfield  {author} {\bibinfo {author} {\bibfnamefont {Y.}~\bibnamefont
  {Tserkovnyak}}, \bibinfo {author} {\bibfnamefont {A.}~\bibnamefont
  {Brataas}}, \bibinfo {author} {\bibfnamefont {G.~E.~W.}\ \bibnamefont
  {Bauer}},\ and\ \bibinfo {author} {\bibfnamefont {B.~I.}\ \bibnamefont
  {Halperin}},\ }\bibfield  {title} {\bibinfo {title} {Nonlocal magnetization
  dynamics in ferromagnetic heterostructures},\ }\href
  {https://doi.org/10.1103/RevModPhys.77.1375} {\bibfield  {journal} {\bibinfo
  {journal} {Rev. Mod. Phys.}\ }\textbf {\bibinfo {volume} {77}},\ \bibinfo
  {pages} {1375} (\bibinfo {year} {2005})}\BibitemShut {NoStop}%
\bibitem [{\citenamefont {Bauer}\ \emph {et~al.}(2012)\citenamefont {Bauer},
  \citenamefont {Saitoh},\ and\ \citenamefont {van Wees}}]{spin-calori-12}%
  \BibitemOpen
  \bibfield  {author} {\bibinfo {author} {\bibfnamefont {G.~E.~W.}\
  \bibnamefont {Bauer}}, \bibinfo {author} {\bibfnamefont {E.}~\bibnamefont
  {Saitoh}},\ and\ \bibinfo {author} {\bibfnamefont {B.~J.}\ \bibnamefont {van
  Wees}},\ }\bibfield  {title} {\bibinfo {title} {Spin caloritronics},\ }\href
  {https://doi.org/10.1038/nmat3301} {\bibfield  {journal} {\bibinfo  {journal}
  {Nat. Mater.}\ }\textbf {\bibinfo {volume} {11}},\ \bibinfo {pages} {391}
  (\bibinfo {year} {2012})}\BibitemShut {NoStop}%
\bibitem [{\citenamefont {Uchida}\ and\ \citenamefont
  {Hirai}(2026)}]{spin-calori-26}%
  \BibitemOpen
  \bibfield  {author} {\bibinfo {author} {\bibfnamefont {K.-i.}\ \bibnamefont
  {Uchida}}\ and\ \bibinfo {author} {\bibfnamefont {T.}~\bibnamefont {Hirai}},\
  }\bibfield  {title} {\bibinfo {title} {Spin caloritronics: history and future
  prospects of experiments},\ }\href@noop {} {\bibfield  {journal} {\bibinfo
  {journal} {J. Magn. Magn. Mater.}\ ,\ \bibinfo {pages} {174195}} (\bibinfo
  {year} {2026})}\BibitemShut {NoStop}%
\bibitem [{\citenamefont {Uchida}\ \emph {et~al.}(2008)\citenamefont {Uchida},
  \citenamefont {Takahashi}, \citenamefont {Harii}, \citenamefont {Ieda},
  \citenamefont {Koshibae}, \citenamefont {Ando}, \citenamefont {Maekawa},\
  and\ \citenamefont {Saitoh}}]{Seebeck08}%
  \BibitemOpen
  \bibfield  {author} {\bibinfo {author} {\bibfnamefont {K.}~\bibnamefont
  {Uchida}}, \bibinfo {author} {\bibfnamefont {S.}~\bibnamefont {Takahashi}},
  \bibinfo {author} {\bibfnamefont {K.}~\bibnamefont {Harii}}, \bibinfo
  {author} {\bibfnamefont {J.}~\bibnamefont {Ieda}}, \bibinfo {author}
  {\bibfnamefont {W.}~\bibnamefont {Koshibae}}, \bibinfo {author}
  {\bibfnamefont {K.}~\bibnamefont {Ando}}, \bibinfo {author} {\bibfnamefont
  {S.}~\bibnamefont {Maekawa}},\ and\ \bibinfo {author} {\bibfnamefont
  {E.}~\bibnamefont {Saitoh}},\ }\bibfield  {title} {\bibinfo {title}
  {Observation of the spin {S}eebeck effect},\ }\href
  {https://doi.org/10.1038/nature07321} {\bibfield  {journal} {\bibinfo
  {journal} {Nature}\ }\textbf {\bibinfo {volume} {455}},\ \bibinfo {pages}
  {778} (\bibinfo {year} {2008})}\BibitemShut {NoStop}%
\bibitem [{\citenamefont {Xiao}\ \emph {et~al.}(2010)\citenamefont {Xiao},
  \citenamefont {Bauer}, \citenamefont {Uchida}, \citenamefont {Saitoh},\ and\
  \citenamefont {Maekawa}}]{Seebeck10}%
  \BibitemOpen
  \bibfield  {author} {\bibinfo {author} {\bibfnamefont {J.}~\bibnamefont
  {Xiao}}, \bibinfo {author} {\bibfnamefont {G.~E.~W.}\ \bibnamefont {Bauer}},
  \bibinfo {author} {\bibfnamefont {K.-c.}\ \bibnamefont {Uchida}}, \bibinfo
  {author} {\bibfnamefont {E.}~\bibnamefont {Saitoh}},\ and\ \bibinfo {author}
  {\bibfnamefont {S.}~\bibnamefont {Maekawa}},\ }\bibfield  {title} {\bibinfo
  {title} {Theory of magnon-driven spin {S}eebeck effect},\ }\href
  {https://doi.org/10.1103/PhysRevB.81.214418} {\bibfield  {journal} {\bibinfo
  {journal} {Phys. Rev. B}\ }\textbf {\bibinfo {volume} {81}},\ \bibinfo
  {pages} {214418} (\bibinfo {year} {2010})}\BibitemShut {NoStop}%
\bibitem [{\citenamefont {Jaworski}\ \emph {et~al.}(2010)\citenamefont
  {Jaworski}, \citenamefont {Yang}, \citenamefont {Mack}, \citenamefont
  {Awschalom}, \citenamefont {Heremans},\ and\ \citenamefont
  {Myers}}]{Seebeck10-1}%
  \BibitemOpen
  \bibfield  {author} {\bibinfo {author} {\bibfnamefont {C.}~\bibnamefont
  {Jaworski}}, \bibinfo {author} {\bibfnamefont {J.}~\bibnamefont {Yang}},
  \bibinfo {author} {\bibfnamefont {S.}~\bibnamefont {Mack}}, \bibinfo {author}
  {\bibfnamefont {D.}~\bibnamefont {Awschalom}}, \bibinfo {author}
  {\bibfnamefont {J.}~\bibnamefont {Heremans}},\ and\ \bibinfo {author}
  {\bibfnamefont {R.}~\bibnamefont {Myers}},\ }\bibfield  {title} {\bibinfo
  {title} {Observation of the spin-{S}eebeck effect in a ferromagnetic
  semiconductor},\ }\href@noop {} {\bibfield  {journal} {\bibinfo  {journal}
  {Nat. Mater.}\ }\textbf {\bibinfo {volume} {9}},\ \bibinfo {pages} {898}
  (\bibinfo {year} {2010})}\BibitemShut {NoStop}%
\bibitem [{\citenamefont {Slachter}\ \emph {et~al.}(2010)\citenamefont
  {Slachter}, \citenamefont {Bakker}, \citenamefont {Adam},\ and\ \citenamefont
  {van Wees}}]{Seebeck10-2}%
  \BibitemOpen
  \bibfield  {author} {\bibinfo {author} {\bibfnamefont {A.}~\bibnamefont
  {Slachter}}, \bibinfo {author} {\bibfnamefont {F.~L.}\ \bibnamefont
  {Bakker}}, \bibinfo {author} {\bibfnamefont {J.-P.}\ \bibnamefont {Adam}},\
  and\ \bibinfo {author} {\bibfnamefont {B.~J.}\ \bibnamefont {van Wees}},\
  }\bibfield  {title} {\bibinfo {title} {Thermally driven spin injection from a
  ferromagnet into a non-magnetic metal},\ }\href
  {https://doi.org/10.1038/nphys1767} {\bibfield  {journal} {\bibinfo
  {journal} {Nat. Phys.}\ }\textbf {\bibinfo {volume} {6}},\ \bibinfo {pages}
  {879} (\bibinfo {year} {2010})}\BibitemShut {NoStop}%
\bibitem [{\citenamefont {Adachi}\ \emph {et~al.}(2013)\citenamefont {Adachi},
  \citenamefont {Uchida}, \citenamefont {Saitoh},\ and\ \citenamefont
  {Maekawa}}]{Seebeck13}%
  \BibitemOpen
  \bibfield  {author} {\bibinfo {author} {\bibfnamefont {H.}~\bibnamefont
  {Adachi}}, \bibinfo {author} {\bibfnamefont {K.-i.}\ \bibnamefont {Uchida}},
  \bibinfo {author} {\bibfnamefont {E.}~\bibnamefont {Saitoh}},\ and\ \bibinfo
  {author} {\bibfnamefont {S.}~\bibnamefont {Maekawa}},\ }\bibfield  {title}
  {\bibinfo {title} {Theory of the spin {S}eebeck effect},\ }\href
  {https://doi.org/10.1088/0034-4885/76/3/036501} {\bibfield  {journal}
  {\bibinfo  {journal} {Rep. Prog. Phys.}\ }\textbf {\bibinfo {volume} {76}},\
  \bibinfo {pages} {036501} (\bibinfo {year} {2013})}\BibitemShut {NoStop}%
\bibitem [{\citenamefont {Tang}\ \emph {et~al.}(2018)\citenamefont {Tang},
  \citenamefont {Chen}, \citenamefont {Ren},\ and\ \citenamefont
  {Wang}}]{Seebeck18}%
  \BibitemOpen
  \bibfield  {author} {\bibinfo {author} {\bibfnamefont {G.}~\bibnamefont
  {Tang}}, \bibinfo {author} {\bibfnamefont {X.}~\bibnamefont {Chen}}, \bibinfo
  {author} {\bibfnamefont {J.}~\bibnamefont {Ren}},\ and\ \bibinfo {author}
  {\bibfnamefont {J.}~\bibnamefont {Wang}},\ }\bibfield  {title} {\bibinfo
  {title} {Rectifying full-counting statistics in a spin {S}eebeck engine},\
  }\href {https://doi.org/10.1103/PhysRevB.97.081407} {\bibfield  {journal}
  {\bibinfo  {journal} {Phys. Rev. B}\ }\textbf {\bibinfo {volume} {97}},\
  \bibinfo {pages} {081407(R)} (\bibinfo {year} {2018})}\BibitemShut {NoStop}%
\bibitem [{\citenamefont {Kim}\ \emph {et~al.}(2023)\citenamefont {Kim},
  \citenamefont {Vetter}, \citenamefont {Yan}, \citenamefont {Yang},
  \citenamefont {Wang}, \citenamefont {Sun}, \citenamefont {Yang},
  \citenamefont {Comstock}, \citenamefont {Li}, \citenamefont {Zhou},
  \citenamefont {Zhang}, \citenamefont {You}, \citenamefont {Sun},\ and\
  \citenamefont {Liu}}]{Seebeck_chiral_23}%
  \BibitemOpen
  \bibfield  {author} {\bibinfo {author} {\bibfnamefont {K.}~\bibnamefont
  {Kim}}, \bibinfo {author} {\bibfnamefont {E.}~\bibnamefont {Vetter}},
  \bibinfo {author} {\bibfnamefont {L.}~\bibnamefont {Yan}}, \bibinfo {author}
  {\bibfnamefont {C.}~\bibnamefont {Yang}}, \bibinfo {author} {\bibfnamefont
  {Z.}~\bibnamefont {Wang}}, \bibinfo {author} {\bibfnamefont {R.}~\bibnamefont
  {Sun}}, \bibinfo {author} {\bibfnamefont {Y.}~\bibnamefont {Yang}}, \bibinfo
  {author} {\bibfnamefont {A.~H.}\ \bibnamefont {Comstock}}, \bibinfo {author}
  {\bibfnamefont {X.}~\bibnamefont {Li}}, \bibinfo {author} {\bibfnamefont
  {J.}~\bibnamefont {Zhou}}, \bibinfo {author} {\bibfnamefont {L.}~\bibnamefont
  {Zhang}}, \bibinfo {author} {\bibfnamefont {W.}~\bibnamefont {You}}, \bibinfo
  {author} {\bibfnamefont {D.}~\bibnamefont {Sun}},\ and\ \bibinfo {author}
  {\bibfnamefont {J.}~\bibnamefont {Liu}},\ }\bibfield  {title} {\bibinfo
  {title} {Chiral-phonon-activated spin {S}eebeck effect},\ }\href
  {https://doi.org/10.1038/s41563-023-01473-9} {\bibfield  {journal} {\bibinfo
  {journal} {Nat. Mater}\ }\textbf {\bibinfo {volume} {22}},\ \bibinfo {pages}
  {322} (\bibinfo {year} {2023})}\BibitemShut {NoStop}%
\bibitem [{\citenamefont {Funato}\ \emph {et~al.}(2024)\citenamefont {Funato},
  \citenamefont {Matsuo},\ and\ \citenamefont {Kato}}]{Seebeck_chiral_24}%
  \BibitemOpen
  \bibfield  {author} {\bibinfo {author} {\bibfnamefont {T.}~\bibnamefont
  {Funato}}, \bibinfo {author} {\bibfnamefont {M.}~\bibnamefont {Matsuo}},\
  and\ \bibinfo {author} {\bibfnamefont {T.}~\bibnamefont {Kato}},\ }\bibfield
  {title} {\bibinfo {title} {Chirality-induced phonon-spin conversion at an
  interface},\ }\href {https://doi.org/10.1103/PhysRevLett.132.236201}
  {\bibfield  {journal} {\bibinfo  {journal} {Phys. Rev. Lett.}\ }\textbf
  {\bibinfo {volume} {132}},\ \bibinfo {pages} {236201} (\bibinfo {year}
  {2024})}\BibitemShut {NoStop}%
\bibitem [{\citenamefont {Nishimura}\ \emph {et~al.}(2025)\citenamefont
  {Nishimura}, \citenamefont {Funato}, \citenamefont {Matsuo},\ and\
  \citenamefont {Kato}}]{Seebeck_chiral_25}%
  \BibitemOpen
  \bibfield  {author} {\bibinfo {author} {\bibfnamefont {N.}~\bibnamefont
  {Nishimura}}, \bibinfo {author} {\bibfnamefont {T.}~\bibnamefont {Funato}},
  \bibinfo {author} {\bibfnamefont {M.}~\bibnamefont {Matsuo}},\ and\ \bibinfo
  {author} {\bibfnamefont {T.}~\bibnamefont {Kato}},\ }\bibfield  {title}
  {\bibinfo {title} {Theory of spin {S}eebeck effect activated by acoustic
  chiral phonons},\ }\href
  {https://doi.org/https://doi.org/10.1016/j.jmmm.2025.173386} {\bibfield
  {journal} {\bibinfo  {journal} {J. Magn. Magn. Mat.}\ }\textbf {\bibinfo
  {volume} {630}},\ \bibinfo {pages} {173386} (\bibinfo {year}
  {2025})}\BibitemShut {NoStop}%
\bibitem [{\citenamefont {Zhang}\ \emph {et~al.}(2026)\citenamefont {Zhang},
  \citenamefont {Li}, \citenamefont {Tang},\ and\ \citenamefont
  {Xing}}]{GT26-4}%
  \BibitemOpen
  \bibfield  {author} {\bibinfo {author} {\bibfnamefont {J.}~\bibnamefont
  {Zhang}}, \bibinfo {author} {\bibfnamefont {G.}~\bibnamefont {Li}}, \bibinfo
  {author} {\bibfnamefont {G.}~\bibnamefont {Tang}},\ and\ \bibinfo {author}
  {\bibfnamefont {Y.}~\bibnamefont {Xing}},\ }\bibfield  {title} {\bibinfo
  {title} {Spin {S}eebeck effect in normal-metal--chiral-insulator
  heterostructures},\ }\href {https://doi.org/10.1103/c2vp-zwtr} {\bibfield
  {journal} {\bibinfo  {journal} {Phys. Rev. B}\ }\textbf {\bibinfo {volume}
  {114}},\ \bibinfo {pages} {L111302} (\bibinfo {year} {2026})}\BibitemShut
  {NoStop}%
\bibitem [{\citenamefont {Dyrda{\l{}}}\ \emph {et~al.}(2018)\citenamefont
  {Dyrda{\l{}}}, \citenamefont {Barna{\ifmmode \acute{s}\else \'{s}\fi{}}},
  \citenamefont {Dugaev},\ and\ \citenamefont {Berakdar}}]{Dyrdal18}%
  \BibitemOpen
  \bibfield  {author} {\bibinfo {author} {\bibfnamefont {A.}~\bibnamefont
  {Dyrda{\l{}}}}, \bibinfo {author} {\bibfnamefont {J.}~\bibnamefont
  {Barna{\ifmmode \acute{s}\else \'{s}\fi{}}}}, \bibinfo {author}
  {\bibfnamefont {V.~K.}\ \bibnamefont {Dugaev}},\ and\ \bibinfo {author}
  {\bibfnamefont {J.}~\bibnamefont {Berakdar}},\ }\bibfield  {title} {\bibinfo
  {title} {Thermally induced spin polarization in a magnetized two-dimensional
  electron gas with {R}ashba spin-orbit interaction},\ }\href
  {https://doi.org/10.1103/PhysRevB.98.075307} {\bibfield  {journal} {\bibinfo
  {journal} {Phys. Rev. B}\ }\textbf {\bibinfo {volume} {98}},\ \bibinfo
  {pages} {075307} (\bibinfo {year} {2018})}\BibitemShut {NoStop}%
\bibitem [{\citenamefont {Meier}\ and\ \citenamefont
  {Zakharchenya}(1984)}]{Meier84}%
  \BibitemOpen
  \bibfield  {author} {\bibinfo {author} {\bibfnamefont {F.}~\bibnamefont
  {Meier}}\ and\ \bibinfo {author} {\bibfnamefont {B.~P.}\ \bibnamefont
  {Zakharchenya}},\ }\href@noop {} {\emph {\bibinfo {title} {Optical
  Orientation}}},\ \bibinfo {series} {Modern Problems in Condensed Matter
  Sciences}, Vol.~\bibinfo {volume} {8}\ (\bibinfo  {publisher} {North-Holland,
  Amsterdam},\ \bibinfo {year} {1984})\BibitemShut {NoStop}%
\bibitem [{\citenamefont {Edelstein}(1998)}]{Edelstein98}%
  \BibitemOpen
  \bibfield  {author} {\bibinfo {author} {\bibfnamefont {V.~M.}\ \bibnamefont
  {Edelstein}},\ }\bibfield  {title} {\bibinfo {title} {Inverse {F}araday
  effect in conducting crystals caused by a broken mirror symmetry},\ }\href
  {https://doi.org/10.1103/PhysRevLett.80.5766} {\bibfield  {journal} {\bibinfo
   {journal} {Phys. Rev. Lett.}\ }\textbf {\bibinfo {volume} {80}},\ \bibinfo
  {pages} {5766} (\bibinfo {year} {1998})}\BibitemShut {NoStop}%
\bibitem [{\citenamefont {Berritta}\ \emph {et~al.}(2016)\citenamefont
  {Berritta}, \citenamefont {Mondal}, \citenamefont {Carva},\ and\
  \citenamefont {Oppeneer}}]{Berritta16}%
  \BibitemOpen
  \bibfield  {author} {\bibinfo {author} {\bibfnamefont {M.}~\bibnamefont
  {Berritta}}, \bibinfo {author} {\bibfnamefont {R.}~\bibnamefont {Mondal}},
  \bibinfo {author} {\bibfnamefont {K.}~\bibnamefont {Carva}},\ and\ \bibinfo
  {author} {\bibfnamefont {P.~M.}\ \bibnamefont {Oppeneer}},\ }\bibfield
  {title} {\bibinfo {title} {Ab initio theory of coherent laser-induced
  magnetization in metals},\ }\href
  {https://doi.org/10.1103/PhysRevLett.117.137203} {\bibfield  {journal}
  {\bibinfo  {journal} {Phys. Rev. Lett.}\ }\textbf {\bibinfo {volume} {117}},\
  \bibinfo {pages} {137203} (\bibinfo {year} {2016})}\BibitemShut {NoStop}%
\bibitem [{\citenamefont {Tanaka}\ and\ \citenamefont {Sato}(2024)}]{Tanaka24}%
  \BibitemOpen
  \bibfield  {author} {\bibinfo {author} {\bibfnamefont {M.}~\bibnamefont
  {Tanaka}}\ and\ \bibinfo {author} {\bibfnamefont {M.}~\bibnamefont {Sato}},\
  }\bibfield  {title} {\bibinfo {title} {Theory of the inverse {F}araday effect
  in dissipative {R}ashba electron systems: {F}loquet engineering
  perspective},\ }\href {https://doi.org/10.1103/PhysRevB.110.045204}
  {\bibfield  {journal} {\bibinfo  {journal} {Phys. Rev. B}\ }\textbf {\bibinfo
  {volume} {110}},\ \bibinfo {pages} {045204} (\bibinfo {year}
  {2024})}\BibitemShut {NoStop}%
\bibitem [{\citenamefont {Yu}\ and\ \citenamefont {Bauer}(2020)}]{YuBauer20}%
  \BibitemOpen
  \bibfield  {author} {\bibinfo {author} {\bibfnamefont {T.}~\bibnamefont
  {Yu}}\ and\ \bibinfo {author} {\bibfnamefont {G.~E.~W.}\ \bibnamefont
  {Bauer}},\ }\bibfield  {title} {\bibinfo {title} {Noncontact spin pumping by
  microwave evanescent fields},\ }\href
  {https://doi.org/10.1103/PhysRevLett.124.236801} {\bibfield  {journal}
  {\bibinfo  {journal} {Phys. Rev. Lett.}\ }\textbf {\bibinfo {volume} {124}},\
  \bibinfo {pages} {236801} (\bibinfo {year} {2020})}\BibitemShut {NoStop}%
\bibitem [{\citenamefont {Li}\ \emph {et~al.}(2026)\citenamefont {Li},
  \citenamefont {Cai},\ and\ \citenamefont {Yu}}]{LiCaiYu26}%
  \BibitemOpen
  \bibfield  {author} {\bibinfo {author} {\bibfnamefont {P.}~\bibnamefont
  {Li}}, \bibinfo {author} {\bibfnamefont {C.}~\bibnamefont {Cai}},\ and\
  \bibinfo {author} {\bibfnamefont {T.}~\bibnamefont {Yu}},\ }\bibfield
  {title} {\bibinfo {title} {Transverse and unidirectional spin pumping},\
  }\href {https://doi.org/10.1103/3l46-tx5b} {\bibfield  {journal} {\bibinfo
  {journal} {Phys. Rev. B}\ }\textbf {\bibinfo {volume} {113}},\ \bibinfo
  {pages} {014438} (\bibinfo {year} {2026})}\BibitemShut {NoStop}%
\bibitem [{\citenamefont {Volokitin}\ and\ \citenamefont
  {Persson}(2007)}]{review07}%
  \BibitemOpen
  \bibfield  {author} {\bibinfo {author} {\bibfnamefont {A.~I.}\ \bibnamefont
  {Volokitin}}\ and\ \bibinfo {author} {\bibfnamefont {B.~N.~J.}\ \bibnamefont
  {Persson}},\ }\bibfield  {title} {\bibinfo {title} {Near-field radiative heat
  transfer and noncontact friction},\ }\href
  {https://doi.org/10.1103/RevModPhys.79.1291} {\bibfield  {journal} {\bibinfo
  {journal} {Rev. Mod. Phys.}\ }\textbf {\bibinfo {volume} {79}},\ \bibinfo
  {pages} {1291} (\bibinfo {year} {2007})}\BibitemShut {NoStop}%
\bibitem [{\citenamefont {Song}\ \emph {et~al.}(2015)\citenamefont {Song},
  \citenamefont {Fiorino}, \citenamefont {Meyhofer},\ and\ \citenamefont
  {Reddy}}]{review15}%
  \BibitemOpen
  \bibfield  {author} {\bibinfo {author} {\bibfnamefont {B.}~\bibnamefont
  {Song}}, \bibinfo {author} {\bibfnamefont {A.}~\bibnamefont {Fiorino}},
  \bibinfo {author} {\bibfnamefont {E.}~\bibnamefont {Meyhofer}},\ and\
  \bibinfo {author} {\bibfnamefont {P.}~\bibnamefont {Reddy}},\ }\bibfield
  {title} {\bibinfo {title} {Near-field radiative thermal transport: From
  theory to experiment},\ }\href {https://doi.org/10.1063/1.4919048} {\bibfield
   {journal} {\bibinfo  {journal} {AIP Adv.}\ }\textbf {\bibinfo {volume}
  {5}},\ \bibinfo {pages} {053503} (\bibinfo {year} {2015})}\BibitemShut
  {NoStop}%
\bibitem [{\citenamefont {Cuevas}\ and\ \citenamefont
  {García-Vidal}(2018)}]{review18}%
  \BibitemOpen
  \bibfield  {author} {\bibinfo {author} {\bibfnamefont {J.~C.}\ \bibnamefont
  {Cuevas}}\ and\ \bibinfo {author} {\bibfnamefont {F.~J.}\ \bibnamefont
  {García-Vidal}},\ }\bibfield  {title} {\bibinfo {title} {Radiative heat
  transfer},\ }\href {https://doi.org/10.1021/acsphotonics.8b01031} {\bibfield
  {journal} {\bibinfo  {journal} {ACS Photonics}\ }\textbf {\bibinfo {volume}
  {5}},\ \bibinfo {pages} {3896} (\bibinfo {year} {2018})}\BibitemShut
  {NoStop}%
\bibitem [{\citenamefont {Biehs}\ \emph {et~al.}(2021)\citenamefont {Biehs},
  \citenamefont {Messina}, \citenamefont {Venkataram}, \citenamefont
  {Rodriguez}, \citenamefont {Cuevas},\ and\ \citenamefont
  {Ben-Abdallah}}]{review21}%
  \BibitemOpen
  \bibfield  {author} {\bibinfo {author} {\bibfnamefont {S.-A.}\ \bibnamefont
  {Biehs}}, \bibinfo {author} {\bibfnamefont {R.}~\bibnamefont {Messina}},
  \bibinfo {author} {\bibfnamefont {P.~S.}\ \bibnamefont {Venkataram}},
  \bibinfo {author} {\bibfnamefont {A.~W.}\ \bibnamefont {Rodriguez}}, \bibinfo
  {author} {\bibfnamefont {J.~C.}\ \bibnamefont {Cuevas}},\ and\ \bibinfo
  {author} {\bibfnamefont {P.}~\bibnamefont {Ben-Abdallah}},\ }\bibfield
  {title} {\bibinfo {title} {Near-field radiative heat transfer in many-body
  systems},\ }\href {https://doi.org/10.1103/RevModPhys.93.025009} {\bibfield
  {journal} {\bibinfo  {journal} {Rev. Mod. Phys.}\ }\textbf {\bibinfo {volume}
  {93}},\ \bibinfo {pages} {025009} (\bibinfo {year} {2021})}\BibitemShut
  {NoStop}%
\bibitem [{\citenamefont {Khandekar}\ and\ \citenamefont
  {Jacob}(2019)}]{Jacob19}%
  \BibitemOpen
  \bibfield  {author} {\bibinfo {author} {\bibfnamefont {C.}~\bibnamefont
  {Khandekar}}\ and\ \bibinfo {author} {\bibfnamefont {Z.}~\bibnamefont
  {Jacob}},\ }\bibfield  {title} {\bibinfo {title} {Thermal spin photonics in
  the near-field of nonreciprocal media},\ }\href
  {https://doi.org/10.1088/1367-2630/ab494d} {\bibfield  {journal} {\bibinfo
  {journal} {New J. Phys.}\ }\textbf {\bibinfo {volume} {21}},\ \bibinfo
  {pages} {103030} (\bibinfo {year} {2019})}\BibitemShut {NoStop}%
\bibitem [{\citenamefont {Gao}\ \emph {et~al.}(2021)\citenamefont {Gao},
  \citenamefont {Khandekar}, \citenamefont {Jacob},\ and\ \citenamefont
  {Li}}]{nonrecip_torque21}%
  \BibitemOpen
  \bibfield  {author} {\bibinfo {author} {\bibfnamefont {X.}~\bibnamefont
  {Gao}}, \bibinfo {author} {\bibfnamefont {C.}~\bibnamefont {Khandekar}},
  \bibinfo {author} {\bibfnamefont {Z.}~\bibnamefont {Jacob}},\ and\ \bibinfo
  {author} {\bibfnamefont {T.}~\bibnamefont {Li}},\ }\bibfield  {title}
  {\bibinfo {title} {Thermal equilibrium spin torque: Near-field radiative
  angular momentum transfer in magneto-optical media},\ }\href
  {https://doi.org/10.1103/PhysRevB.103.125424} {\bibfield  {journal} {\bibinfo
   {journal} {Phys. Rev. B}\ }\textbf {\bibinfo {volume} {103}},\ \bibinfo
  {pages} {125424} (\bibinfo {year} {2021})}\BibitemShut {NoStop}%
\bibitem [{\citenamefont {Khan}\ and\ \citenamefont
  {Narimanov}(2019)}]{Khan19}%
  \BibitemOpen
  \bibfield  {author} {\bibinfo {author} {\bibfnamefont {E.}~\bibnamefont
  {Khan}}\ and\ \bibinfo {author} {\bibfnamefont {E.~E.}\ \bibnamefont
  {Narimanov}},\ }\bibfield  {title} {\bibinfo {title} {Spinning radiation from
  a topological insulator},\ }\href
  {https://doi.org/10.1103/PhysRevB.100.081408} {\bibfield  {journal} {\bibinfo
   {journal} {Phys. Rev. B}\ }\textbf {\bibinfo {volume} {100}},\ \bibinfo
  {pages} {081408(R)} (\bibinfo {year} {2019})}\BibitemShut {NoStop}%
\bibitem [{\citenamefont {Zhu}\ \emph {et~al.}(2024)\citenamefont {Zhu},
  \citenamefont {Tang}, \citenamefont {Zhang},\ and\ \citenamefont
  {Chen}}]{GT24-2}%
  \BibitemOpen
  \bibfield  {author} {\bibinfo {author} {\bibfnamefont {H.}~\bibnamefont
  {Zhu}}, \bibinfo {author} {\bibfnamefont {G.}~\bibnamefont {Tang}}, \bibinfo
  {author} {\bibfnamefont {L.}~\bibnamefont {Zhang}},\ and\ \bibinfo {author}
  {\bibfnamefont {J.}~\bibnamefont {Chen}},\ }\bibfield  {title} {\bibinfo
  {title} {Current-induced near-field radiative energy, linear-momentum, and
  angular-momentum transfer},\ }\href
  {https://doi.org/10.1103/PhysRevB.109.075413} {\bibfield  {journal} {\bibinfo
   {journal} {Phys. Rev. B}\ }\textbf {\bibinfo {volume} {109}},\ \bibinfo
  {pages} {075413} (\bibinfo {year} {2024})}\BibitemShut {NoStop}%
\bibitem [{\citenamefont {Ben-Abdallah}(2026)}]{spin-PBA-26}%
  \BibitemOpen
  \bibfield  {author} {\bibinfo {author} {\bibfnamefont {P.}~\bibnamefont
  {Ben-Abdallah}},\ }\bibfield  {title} {\bibinfo {title} {Radiative spin
  caloritronics in nonreciprocal many-body systems},\ }\href
  {https://doi.org/10.1103/zxkf-9269} {\bibfield  {journal} {\bibinfo
  {journal} {Phys. Rev. B}\ }\textbf {\bibinfo {volume} {114}},\ \bibinfo
  {pages} {285402} (\bibinfo {year} {2026})}\BibitemShut {NoStop}%
\bibitem [{\citenamefont {He}\ and\ \citenamefont {Tang}(2026)}]{GT26-2}%
  \BibitemOpen
  \bibfield  {author} {\bibinfo {author} {\bibfnamefont {D.}~\bibnamefont
  {He}}\ and\ \bibinfo {author} {\bibfnamefont {G.}~\bibnamefont {Tang}},\
  }\bibfield  {title} {\bibinfo {title} {Transverse thermophotovoltaics from
  nonreciprocal plasmon drag in metal},\ }\href
  {https://doi.org/10.1103/kl16-bw43} {\bibfield  {journal} {\bibinfo
  {journal} {Phys. Rev. Lett.}\ }\textbf {\bibinfo {volume} {136}},\ \bibinfo
  {pages} {176901} (\bibinfo {year} {2026})}\BibitemShut {NoStop}%
\bibitem [{SM()}]{SM}%
  \BibitemOpen
  \href@noop {} {\bibinfo {title} {{See Supplemental Material for detailed
  derivations of the spin current using the nonequilibrium Green's function
  formalism, evaluations of the chiral and nonchiral responses, and the
  connection between the dispersive response and the Zeeman quantum geometric
  tensor in the clean limit. The Supplemental Material also includes
  Refs.~\cite{GT26-3,JW25-1, JW25-2, JW26}.}}}\BibitemShut {Stop}%
\bibitem [{\citenamefont {Hofmann}\ and\ \citenamefont
  {Das~Sarma}(2016)}]{WSM_SPP16}%
  \BibitemOpen
  \bibfield  {author} {\bibinfo {author} {\bibfnamefont {J.}~\bibnamefont
  {Hofmann}}\ and\ \bibinfo {author} {\bibfnamefont {S.}~\bibnamefont
  {Das~Sarma}},\ }\bibfield  {title} {\bibinfo {title} {Surface plasmon
  polaritons in topological {W}eyl semimetals},\ }\href
  {https://doi.org/10.1103/PhysRevB.93.241402} {\bibfield  {journal} {\bibinfo
  {journal} {Phys. Rev. B}\ }\textbf {\bibinfo {volume} {93}},\ \bibinfo
  {pages} {241402(R)} (\bibinfo {year} {2016})}\BibitemShut {NoStop}%
\bibitem [{\citenamefont {Kotov}\ and\ \citenamefont
  {Lozovik}(2018)}]{Kotov18}%
  \BibitemOpen
  \bibfield  {author} {\bibinfo {author} {\bibfnamefont {O.~V.}\ \bibnamefont
  {Kotov}}\ and\ \bibinfo {author} {\bibfnamefont {Y.~E.}\ \bibnamefont
  {Lozovik}},\ }\bibfield  {title} {\bibinfo {title} {Giant tunable
  nonreciprocity of light in {W}eyl semimetals},\ }\href
  {https://doi.org/10.1103/PhysRevB.98.195446} {\bibfield  {journal} {\bibinfo
  {journal} {Phys. Rev. B}\ }\textbf {\bibinfo {volume} {98}},\ \bibinfo
  {pages} {195446} (\bibinfo {year} {2018})}\BibitemShut {NoStop}%
\bibitem [{\citenamefont {Li}\ \emph {et~al.}(2020)\citenamefont {Li},
  \citenamefont {Koo}, \citenamefont {Ning}, \citenamefont {Li}, \citenamefont
  {Miao}, \citenamefont {Min}, \citenamefont {Zhu}, \citenamefont {Wang},
  \citenamefont {Alem}, \citenamefont {Liu}, \citenamefont {Mao},\ and\
  \citenamefont {Yan}}]{WSM_AHE_20nc}%
  \BibitemOpen
  \bibfield  {author} {\bibinfo {author} {\bibfnamefont {P.}~\bibnamefont
  {Li}}, \bibinfo {author} {\bibfnamefont {J.}~\bibnamefont {Koo}}, \bibinfo
  {author} {\bibfnamefont {W.}~\bibnamefont {Ning}}, \bibinfo {author}
  {\bibfnamefont {J.}~\bibnamefont {Li}}, \bibinfo {author} {\bibfnamefont
  {L.}~\bibnamefont {Miao}}, \bibinfo {author} {\bibfnamefont {L.}~\bibnamefont
  {Min}}, \bibinfo {author} {\bibfnamefont {Y.}~\bibnamefont {Zhu}}, \bibinfo
  {author} {\bibfnamefont {Y.}~\bibnamefont {Wang}}, \bibinfo {author}
  {\bibfnamefont {N.}~\bibnamefont {Alem}}, \bibinfo {author} {\bibfnamefont
  {C.-X.}\ \bibnamefont {Liu}}, \bibinfo {author} {\bibfnamefont
  {Z.}~\bibnamefont {Mao}},\ and\ \bibinfo {author} {\bibfnamefont
  {B.}~\bibnamefont {Yan}},\ }\bibfield  {title} {\bibinfo {title} {Giant room
  temperature anomalous {H}all effect and tunable topology in a ferromagnetic
  topological semimetal {C}o$_2${M}n{A}l},\ }\href
  {https://doi.org/10.1038/s41467-020-17174-9} {\bibfield  {journal} {\bibinfo
  {journal} {Nat. Commun.}\ }\textbf {\bibinfo {volume} {11}},\ \bibinfo
  {pages} {3476} (\bibinfo {year} {2020})}\BibitemShut {NoStop}%
\bibitem [{\citenamefont {Tang}\ \emph {et~al.}(2021)\citenamefont {Tang},
  \citenamefont {Chen},\ and\ \citenamefont {Zhang}}]{GT_WSM}%
  \BibitemOpen
  \bibfield  {author} {\bibinfo {author} {\bibfnamefont {G.}~\bibnamefont
  {Tang}}, \bibinfo {author} {\bibfnamefont {J.}~\bibnamefont {Chen}},\ and\
  \bibinfo {author} {\bibfnamefont {L.}~\bibnamefont {Zhang}},\ }\bibfield
  {title} {\bibinfo {title} {Twist-induced control of near-field heat radiation
  between magnetic {W}eyl semimetals},\ }\href
  {https://doi.org/10.1021/acsphotonics.0c01945} {\bibfield  {journal}
  {\bibinfo  {journal} {ACS Photonics}\ }\textbf {\bibinfo {volume} {8}},\
  \bibinfo {pages} {443} (\bibinfo {year} {2021})}\BibitemShut {NoStop}%
\bibitem [{\citenamefont {Park}\ \emph {et~al.}(2026)\citenamefont {Park},
  \citenamefont {Zhang}, \citenamefont {Cheong}, \citenamefont {Kim},
  \citenamefont {Belvin}, \citenamefont {Hsieh}, \citenamefont {Ning},\ and\
  \citenamefont {Gedik}}]{2D_magnets_26}%
  \BibitemOpen
  \bibfield  {author} {\bibinfo {author} {\bibfnamefont {J.-G.}\ \bibnamefont
  {Park}}, \bibinfo {author} {\bibfnamefont {K.-X.}\ \bibnamefont {Zhang}},
  \bibinfo {author} {\bibfnamefont {H.}~\bibnamefont {Cheong}}, \bibinfo
  {author} {\bibfnamefont {J.~H.}\ \bibnamefont {Kim}}, \bibinfo {author}
  {\bibfnamefont {C.~A.}\ \bibnamefont {Belvin}}, \bibinfo {author}
  {\bibfnamefont {D.}~\bibnamefont {Hsieh}}, \bibinfo {author} {\bibfnamefont
  {H.}~\bibnamefont {Ning}},\ and\ \bibinfo {author} {\bibfnamefont
  {N.}~\bibnamefont {Gedik}},\ }\bibfield  {title} {\bibinfo {title} {2{D} van
  der {W}aals magnets: {F}rom fundamental physics to applications},\ }\href
  {https://doi.org/10.1103/2pff-xy6n} {\bibfield  {journal} {\bibinfo
  {journal} {Rev. Mod. Phys.}\ }\textbf {\bibinfo {volume} {98}},\ \bibinfo
  {pages} {025003} (\bibinfo {year} {2026})}\BibitemShut {NoStop}%
\bibitem [{\citenamefont {Tserkovnyak}\ \emph {et~al.}(2002)\citenamefont
  {Tserkovnyak}, \citenamefont {Brataas},\ and\ \citenamefont {Bauer}}]{SP02}%
  \BibitemOpen
  \bibfield  {author} {\bibinfo {author} {\bibfnamefont {Y.}~\bibnamefont
  {Tserkovnyak}}, \bibinfo {author} {\bibfnamefont {A.}~\bibnamefont
  {Brataas}},\ and\ \bibinfo {author} {\bibfnamefont {G.~E.~W.}\ \bibnamefont
  {Bauer}},\ }\bibfield  {title} {\bibinfo {title} {Enhanced {G}ilbert damping
  in thin ferromagnetic films},\ }\href
  {https://doi.org/10.1103/PhysRevLett.88.117601} {\bibfield  {journal}
  {\bibinfo  {journal} {Phys. Rev. Lett.}\ }\textbf {\bibinfo {volume} {88}},\
  \bibinfo {pages} {117601} (\bibinfo {year} {2002})}\BibitemShut {NoStop}%
\bibitem [{\citenamefont {Ominato}\ \emph {et~al.}(2025)\citenamefont
  {Ominato}, \citenamefont {Yama}, \citenamefont {Yamakage}, \citenamefont
  {Matsuo},\ and\ \citenamefont {Kato}}]{SP25}%
  \BibitemOpen
  \bibfield  {author} {\bibinfo {author} {\bibfnamefont {Y.}~\bibnamefont
  {Ominato}}, \bibinfo {author} {\bibfnamefont {M.}~\bibnamefont {Yama}},
  \bibinfo {author} {\bibfnamefont {A.}~\bibnamefont {Yamakage}}, \bibinfo
  {author} {\bibfnamefont {M.}~\bibnamefont {Matsuo}},\ and\ \bibinfo {author}
  {\bibfnamefont {T.}~\bibnamefont {Kato}},\ }\bibfield  {title} {\bibinfo
  {title} {Spin pumping into two-dimensional systems},\ }\href@noop {}
  {\bibfield  {journal} {\bibinfo  {journal} {J. Phys. Condens. Matter}\
  }\textbf {\bibinfo {volume} {37}},\ \bibinfo {pages} {433001} (\bibinfo
  {year} {2025})}\BibitemShut {NoStop}%
\bibitem [{\citenamefont {Li}\ \emph {et~al.}(2019)\citenamefont {Li},
  \citenamefont {Fern\'andez-Alc\'azar}, \citenamefont {Ellis}, \citenamefont
  {Shapiro},\ and\ \citenamefont {Kottos}}]{Li19}%
  \BibitemOpen
  \bibfield  {author} {\bibinfo {author} {\bibfnamefont {H.}~\bibnamefont
  {Li}}, \bibinfo {author} {\bibfnamefont {L.~J.}\ \bibnamefont
  {Fern\'andez-Alc\'azar}}, \bibinfo {author} {\bibfnamefont {F.}~\bibnamefont
  {Ellis}}, \bibinfo {author} {\bibfnamefont {B.}~\bibnamefont {Shapiro}},\
  and\ \bibinfo {author} {\bibfnamefont {T.}~\bibnamefont {Kottos}},\
  }\bibfield  {title} {\bibinfo {title} {Adiabatic thermal radiation pumps for
  thermal photonics},\ }\href {https://doi.org/10.1103/PhysRevLett.123.165901}
  {\bibfield  {journal} {\bibinfo  {journal} {Phys. Rev. Lett.}\ }\textbf
  {\bibinfo {volume} {123}},\ \bibinfo {pages} {165901} (\bibinfo {year}
  {2019})}\BibitemShut {NoStop}%
\bibitem [{\citenamefont {Fern\'andez-Alc\'azar}\ \emph
  {et~al.}(2021)\citenamefont {Fern\'andez-Alc\'azar}, \citenamefont
  {Kononchuk}, \citenamefont {Li},\ and\ \citenamefont {Kottos}}]{Li21}%
  \BibitemOpen
  \bibfield  {author} {\bibinfo {author} {\bibfnamefont {L.~J.}\ \bibnamefont
  {Fern\'andez-Alc\'azar}}, \bibinfo {author} {\bibfnamefont {R.}~\bibnamefont
  {Kononchuk}}, \bibinfo {author} {\bibfnamefont {H.}~\bibnamefont {Li}},\ and\
  \bibinfo {author} {\bibfnamefont {T.}~\bibnamefont {Kottos}},\ }\bibfield
  {title} {\bibinfo {title} {Extreme nonreciprocal near-field thermal radiation
  via {F}loquet photonics},\ }\href
  {https://doi.org/10.1103/PhysRevLett.126.204101} {\bibfield  {journal}
  {\bibinfo  {journal} {Phys. Rev. Lett.}\ }\textbf {\bibinfo {volume} {126}},\
  \bibinfo {pages} {204101} (\bibinfo {year} {2021})}\BibitemShut {NoStop}%
\bibitem [{\citenamefont {Biehs}\ and\ \citenamefont
  {Agarwal}(2023)}]{Biehs22}%
  \BibitemOpen
  \bibfield  {author} {\bibinfo {author} {\bibfnamefont {S.-A.}\ \bibnamefont
  {Biehs}}\ and\ \bibinfo {author} {\bibfnamefont {G.~S.}\ \bibnamefont
  {Agarwal}},\ }\bibfield  {title} {\bibinfo {title} {Breakdown of detailed
  balance for thermal radiation by synthetic fields},\ }\href
  {https://doi.org/10.1103/PhysRevLett.130.110401} {\bibfield  {journal}
  {\bibinfo  {journal} {Phys. Rev. Lett.}\ }\textbf {\bibinfo {volume} {130}},\
  \bibinfo {pages} {110401} (\bibinfo {year} {2023})}\BibitemShut {NoStop}%
\bibitem [{\citenamefont {Tang}\ and\ \citenamefont {Wang}(2024)}]{GT24}%
  \BibitemOpen
  \bibfield  {author} {\bibinfo {author} {\bibfnamefont {G.}~\bibnamefont
  {Tang}}\ and\ \bibinfo {author} {\bibfnamefont {J.-S.}\ \bibnamefont
  {Wang}},\ }\bibfield  {title} {\bibinfo {title} {Modulating near-field
  thermal transfer through temporal drivings: {A} quantum many-body theory},\
  }\href {https://doi.org/10.1103/PhysRevB.109.085428} {\bibfield  {journal}
  {\bibinfo  {journal} {Phys. Rev. B}\ }\textbf {\bibinfo {volume} {109}},\
  \bibinfo {pages} {085428} (\bibinfo {year} {2024})}\BibitemShut {NoStop}%
\bibitem [{\citenamefont {Yu}\ and\ \citenamefont {Fan}(2024)}]{YuFan24}%
  \BibitemOpen
  \bibfield  {author} {\bibinfo {author} {\bibfnamefont {R.}~\bibnamefont
  {Yu}}\ and\ \bibinfo {author} {\bibfnamefont {S.}~\bibnamefont {Fan}},\
  }\bibfield  {title} {\bibinfo {title} {Time-modulated near-field radiative
  heat transfer},\ }\href {https://doi.org/10.1073/pnas.2401514121} {\bibfield
  {journal} {\bibinfo  {journal} {Proc. Natl. Acad. Sci. U.S.A.}\ }\textbf
  {\bibinfo {volume} {121}},\ \bibinfo {pages} {e2401514121} (\bibinfo {year}
  {2024})}\BibitemShut {NoStop}%
\bibitem [{\citenamefont {Pan}\ \emph {et~al.}(2025)\citenamefont {Pan},
  \citenamefont {Ren}, \citenamefont {Tang},\ and\ \citenamefont
  {Wang}}]{GT25}%
  \BibitemOpen
  \bibfield  {author} {\bibinfo {author} {\bibfnamefont {H.}~\bibnamefont
  {Pan}}, \bibinfo {author} {\bibfnamefont {Y.}~\bibnamefont {Ren}}, \bibinfo
  {author} {\bibfnamefont {G.}~\bibnamefont {Tang}},\ and\ \bibinfo {author}
  {\bibfnamefont {J.-S.}\ \bibnamefont {Wang}},\ }\bibfield  {title} {\bibinfo
  {title} {Asymmetry-induced radiative heat transfer in {F}loquet systems},\
  }\href {https://doi.org/10.1103/74rq-f642} {\bibfield  {journal} {\bibinfo
  {journal} {Phys. Rev. B}\ }\textbf {\bibinfo {volume} {112}},\ \bibinfo
  {pages} {L041401} (\bibinfo {year} {2025})}\BibitemShut {NoStop}%
\bibitem [{\citenamefont {Lifshitz}\ and\ \citenamefont
  {Pitaevskii}(2013)}]{Lifshitz_book}%
  \BibitemOpen
  \bibfield  {author} {\bibinfo {author} {\bibfnamefont {E.~M.}\ \bibnamefont
  {Lifshitz}}\ and\ \bibinfo {author} {\bibfnamefont {L.~P.}\ \bibnamefont
  {Pitaevskii}},\ }\href@noop {} {\emph {\bibinfo {title} {Statistical Physics:
  Theory of the Condensed State}}},\ Vol.~\bibinfo {volume} {9}\ (\bibinfo
  {publisher} {Elsevier},\ \bibinfo {year} {2013})\BibitemShut {NoStop}%
\bibitem [{\citenamefont {Haug}\ and\ \citenamefont
  {Jauho}(2008)}]{Haug_Jauho}%
  \BibitemOpen
  \bibfield  {author} {\bibinfo {author} {\bibfnamefont {H.}~\bibnamefont
  {Haug}}\ and\ \bibinfo {author} {\bibfnamefont {A.-P.}\ \bibnamefont
  {Jauho}},\ }\href@noop {} {\emph {\bibinfo {title} {Quantum kinetics in
  transport and optics of semiconductors}}},\ Vol.~\bibinfo {volume} {2}\
  (\bibinfo  {publisher} {Springer},\ \bibinfo {year} {2008})\BibitemShut
  {NoStop}%
\bibitem [{\citenamefont {Wang}\ \emph {et~al.}(2023)\citenamefont {Wang},
  \citenamefont {Peng}, \citenamefont {Zhang}, \citenamefont {Zhang},\ and\
  \citenamefont {Zhu}}]{JSW23}%
  \BibitemOpen
  \bibfield  {author} {\bibinfo {author} {\bibfnamefont {J.-S.}\ \bibnamefont
  {Wang}}, \bibinfo {author} {\bibfnamefont {J.}~\bibnamefont {Peng}}, \bibinfo
  {author} {\bibfnamefont {Z.-Q.}\ \bibnamefont {Zhang}}, \bibinfo {author}
  {\bibfnamefont {Y.-M.}\ \bibnamefont {Zhang}},\ and\ \bibinfo {author}
  {\bibfnamefont {T.}~\bibnamefont {Zhu}},\ }\bibfield  {title} {\bibinfo
  {title} {Transport in electron-photon systems},\ }\href@noop {} {\bibfield
  {journal} {\bibinfo  {journal} {Front. Phys.}\ }\textbf {\bibinfo {volume}
  {18}},\ \bibinfo {pages} {43602} (\bibinfo {year} {2023})}\BibitemShut
  {NoStop}%
\bibitem [{\citenamefont {Wang}(2025)}]{JSW25}%
  \BibitemOpen
  \bibfield  {author} {\bibinfo {author} {\bibfnamefont {J.-S.}\ \bibnamefont
  {Wang}},\ }\bibfield  {title} {\bibinfo {title} {Beyond the {D}rude model:
  Surface and nonlocal effects in near-field radiative heat transfer and the
  {C}asimir puzzle},\ }\href {https://doi.org/10.1103/PhysRevB.111.245404}
  {\bibfield  {journal} {\bibinfo  {journal} {Phys. Rev. B}\ }\textbf {\bibinfo
  {volume} {111}},\ \bibinfo {pages} {245404} (\bibinfo {year}
  {2025})}\BibitemShut {NoStop}%
\bibitem [{\citenamefont {D'yakonov}\ and\ \citenamefont
  {Perel'}(1972)}]{DP72}%
  \BibitemOpen
  \bibfield  {author} {\bibinfo {author} {\bibfnamefont {M.~I.}\ \bibnamefont
  {D'yakonov}}\ and\ \bibinfo {author} {\bibfnamefont {V.~I.}\ \bibnamefont
  {Perel'}},\ }\bibfield  {title} {\bibinfo {title} {Spin relaxation of
  conduction electrons in noncentrosymmetric semiconductors},\ }\href@noop {}
  {\bibfield  {journal} {\bibinfo  {journal} {Sov. Phys. Solid State}\ }\textbf
  {\bibinfo {volume} {13}},\ \bibinfo {pages} {3023} (\bibinfo {year}
  {1972})}\BibitemShut {NoStop}%
\bibitem [{\citenamefont {Ochoa}\ and\ \citenamefont
  {Rold{\'a}n}(2013)}]{DP13}%
  \BibitemOpen
  \bibfield  {author} {\bibinfo {author} {\bibfnamefont {H.}~\bibnamefont
  {Ochoa}}\ and\ \bibinfo {author} {\bibfnamefont {R.}~\bibnamefont
  {Rold{\'a}n}},\ }\bibfield  {title} {\bibinfo {title} {Spin-orbit-mediated
  spin relaxation in monolayer {MoS}$_2$},\ }\href
  {https://doi.org/10.1103/PhysRevB.87.245421} {\bibfield  {journal} {\bibinfo
  {journal} {Phys. Rev. B}\ }\textbf {\bibinfo {volume} {87}},\ \bibinfo
  {pages} {245421} (\bibinfo {year} {2013})}\BibitemShut {NoStop}%
\bibitem [{\citenamefont {Cummings}\ \emph {et~al.}(2017)\citenamefont
  {Cummings}, \citenamefont {Garcia}, \citenamefont {Fabian},\ and\
  \citenamefont {Roche}}]{DP17}%
  \BibitemOpen
  \bibfield  {author} {\bibinfo {author} {\bibfnamefont {A.~W.}\ \bibnamefont
  {Cummings}}, \bibinfo {author} {\bibfnamefont {J.~H.}\ \bibnamefont
  {Garcia}}, \bibinfo {author} {\bibfnamefont {J.}~\bibnamefont {Fabian}},\
  and\ \bibinfo {author} {\bibfnamefont {S.}~\bibnamefont {Roche}},\ }\bibfield
   {title} {\bibinfo {title} {Giant spin lifetime anisotropy in graphene
  induced by proximity effects},\ }\href
  {https://doi.org/10.1103/PhysRevLett.119.206601} {\bibfield  {journal}
  {\bibinfo  {journal} {Phys. Rev. Lett.}\ }\textbf {\bibinfo {volume} {119}},\
  \bibinfo {pages} {206601} (\bibinfo {year} {2017})}\BibitemShut {NoStop}%
\bibitem [{\citenamefont {Yang}\ \emph {et~al.}(2015)\citenamefont {Yang},
  \citenamefont {Sinitsyn}, \citenamefont {Chen}, \citenamefont {Yuan},
  \citenamefont {Zhang}, \citenamefont {Lou},\ and\ \citenamefont
  {Crooker}}]{Yang15}%
  \BibitemOpen
  \bibfield  {author} {\bibinfo {author} {\bibfnamefont {L.}~\bibnamefont
  {Yang}}, \bibinfo {author} {\bibfnamefont {N.~A.}\ \bibnamefont {Sinitsyn}},
  \bibinfo {author} {\bibfnamefont {W.}~\bibnamefont {Chen}}, \bibinfo {author}
  {\bibfnamefont {J.}~\bibnamefont {Yuan}}, \bibinfo {author} {\bibfnamefont
  {J.}~\bibnamefont {Zhang}}, \bibinfo {author} {\bibfnamefont
  {J.}~\bibnamefont {Lou}},\ and\ \bibinfo {author} {\bibfnamefont
  {S.}~\bibnamefont {Crooker}},\ }\bibfield  {title} {\bibinfo {title}
  {Long-lived nanosecond spin relaxation and spin coherence of electrons in
  monolayer {MoS$_2$} and {WS$_2$}},\ }\href
  {https://doi.org/10.1038/nphys3419} {\bibfield  {journal} {\bibinfo
  {journal} {Nature Physics}\ }\textbf {\bibinfo {volume} {11}},\ \bibinfo
  {pages} {830} (\bibinfo {year} {2015})}\BibitemShut {NoStop}%
\bibitem [{\citenamefont {Dey}\ \emph {et~al.}(2017)\citenamefont {Dey},
  \citenamefont {Yang}, \citenamefont {Robert}, \citenamefont {Wang},
  \citenamefont {Urbaszek}, \citenamefont {Marie},\ and\ \citenamefont
  {Crooker}}]{Dey17}%
  \BibitemOpen
  \bibfield  {author} {\bibinfo {author} {\bibfnamefont {P.}~\bibnamefont
  {Dey}}, \bibinfo {author} {\bibfnamefont {L.}~\bibnamefont {Yang}}, \bibinfo
  {author} {\bibfnamefont {C.}~\bibnamefont {Robert}}, \bibinfo {author}
  {\bibfnamefont {G.}~\bibnamefont {Wang}}, \bibinfo {author} {\bibfnamefont
  {B.}~\bibnamefont {Urbaszek}}, \bibinfo {author} {\bibfnamefont
  {X.}~\bibnamefont {Marie}},\ and\ \bibinfo {author} {\bibfnamefont {S.~A.}\
  \bibnamefont {Crooker}},\ }\bibfield  {title} {\bibinfo {title}
  {Gate-controlled spin-valley locking of resident carriers in {WSe$_2$}
  monolayers},\ }\href {https://doi.org/10.1103/PhysRevLett.119.137401}
  {\bibfield  {journal} {\bibinfo  {journal} {Phys. Rev. Lett.}\ }\textbf
  {\bibinfo {volume} {119}},\ \bibinfo {pages} {137401} (\bibinfo {year}
  {2017})}\BibitemShut {NoStop}%
\bibitem [{\citenamefont {Back}\ \emph {et~al.}(2017)\citenamefont {Back},
  \citenamefont {Sidler}, \citenamefont {Cotlet}, \citenamefont {Srivastava},
  \citenamefont {Takemura}, \citenamefont {Kroner},\ and\ \citenamefont
  {Imamo{\ifmmode \breve{g}\else \u{g}\fi{}}lu}}]{Back17}%
  \BibitemOpen
  \bibfield  {author} {\bibinfo {author} {\bibfnamefont {P.}~\bibnamefont
  {Back}}, \bibinfo {author} {\bibfnamefont {M.}~\bibnamefont {Sidler}},
  \bibinfo {author} {\bibfnamefont {O.}~\bibnamefont {Cotlet}}, \bibinfo
  {author} {\bibfnamefont {A.}~\bibnamefont {Srivastava}}, \bibinfo {author}
  {\bibfnamefont {N.}~\bibnamefont {Takemura}}, \bibinfo {author}
  {\bibfnamefont {M.}~\bibnamefont {Kroner}},\ and\ \bibinfo {author}
  {\bibfnamefont {A.}~\bibnamefont {Imamo{\ifmmode \breve{g}\else
  \u{g}\fi{}}lu}},\ }\bibfield  {title} {\bibinfo {title} {Giant
  paramagnetism-induced valley polarization of electrons in charge-tunable
  monolayer {MoSe$_2$}},\ }\href
  {https://doi.org/10.1103/PhysRevLett.118.237404} {\bibfield  {journal}
  {\bibinfo  {journal} {Phys. Rev. Lett.}\ }\textbf {\bibinfo {volume} {118}},\
  \bibinfo {pages} {237404} (\bibinfo {year} {2017})}\BibitemShut {NoStop}%
\bibitem [{\citenamefont {Tang}\ \emph {et~al.}(2026)\citenamefont {Tang},
  \citenamefont {Ren},\ and\ \citenamefont {Wang}}]{GT26-3}%
  \BibitemOpen
  \bibfield  {author} {\bibinfo {author} {\bibfnamefont {G.}~\bibnamefont
  {Tang}}, \bibinfo {author} {\bibfnamefont {Y.}~\bibnamefont {Ren}},\ and\
  \bibinfo {author} {\bibfnamefont {J.-S.}\ \bibnamefont {Wang}},\ }\bibfield
  {title} {\bibinfo {title} {Infinitesimal environmental dust as a photonic
  bath at infinity},\ }\href {https://doi.org/10.1103/855g-71dh} {\bibfield
  {journal} {\bibinfo  {journal} {Phys. Rev. A}\ }\textbf {\bibinfo {volume}
  {114}},\ \bibinfo {pages} {023517} (\bibinfo {year} {2026})}\BibitemShut
  {NoStop}%
\bibitem [{\citenamefont {Xiang}\ \emph
  {et~al.}(2025{\natexlab{a}})\citenamefont {Xiang}, \citenamefont {Jia},
  \citenamefont {Xu}, \citenamefont {Qiao},\ and\ \citenamefont
  {Wang}}]{JW25-1}%
  \BibitemOpen
  \bibfield  {author} {\bibinfo {author} {\bibfnamefont {L.}~\bibnamefont
  {Xiang}}, \bibinfo {author} {\bibfnamefont {J.}~\bibnamefont {Jia}}, \bibinfo
  {author} {\bibfnamefont {F.}~\bibnamefont {Xu}}, \bibinfo {author}
  {\bibfnamefont {Z.}~\bibnamefont {Qiao}},\ and\ \bibinfo {author}
  {\bibfnamefont {J.}~\bibnamefont {Wang}},\ }\bibfield  {title} {\bibinfo
  {title} {Intrinsic gyrotropic magnetic current from {Z}eeman quantum
  geometry},\ }\href {https://doi.org/10.1103/PhysRevLett.134.116301}
  {\bibfield  {journal} {\bibinfo  {journal} {Phys. Rev. Lett.}\ }\textbf
  {\bibinfo {volume} {134}},\ \bibinfo {pages} {116301} (\bibinfo {year}
  {2025}{\natexlab{a}})}\BibitemShut {NoStop}%
\bibitem [{\citenamefont {Xiang}\ \emph
  {et~al.}(2025{\natexlab{b}})\citenamefont {Xiang}, \citenamefont {Jin},\ and\
  \citenamefont {Wang}}]{JW25-2}%
  \BibitemOpen
  \bibfield  {author} {\bibinfo {author} {\bibfnamefont {L.}~\bibnamefont
  {Xiang}}, \bibinfo {author} {\bibfnamefont {H.}~\bibnamefont {Jin}},\ and\
  \bibinfo {author} {\bibfnamefont {J.}~\bibnamefont {Wang}},\ }\bibfield
  {title} {\bibinfo {title} {Spin transport revealed by spin quantum
  geometry},\ }\href {https://doi.org/10.1103/14mp-263q} {\bibfield  {journal}
  {\bibinfo  {journal} {Phys. Rev. Lett.}\ }\textbf {\bibinfo {volume} {135}},\
  \bibinfo {pages} {146303} (\bibinfo {year} {2025}{\natexlab{b}})}\BibitemShut
  {NoStop}%
\bibitem [{\citenamefont {Xiang}\ \emph {et~al.}(2026)\citenamefont {Xiang},
  \citenamefont {Jia}, \citenamefont {Xu},\ and\ \citenamefont {Wang}}]{JW26}%
  \BibitemOpen
  \bibfield  {author} {\bibinfo {author} {\bibfnamefont {L.}~\bibnamefont
  {Xiang}}, \bibinfo {author} {\bibfnamefont {J.}~\bibnamefont {Jia}}, \bibinfo
  {author} {\bibfnamefont {F.}~\bibnamefont {Xu}},\ and\ \bibinfo {author}
  {\bibfnamefont {J.}~\bibnamefont {Wang}},\ }\bibfield  {title} {\bibinfo
  {title} {Quantum geometric map of magnetotransport},\ }\href
  {https://doi.org/10.1103/kd12-4ycj} {\bibfield  {journal} {\bibinfo
  {journal} {Phys. Rev. B}\ }\textbf {\bibinfo {volume} {113}},\ \bibinfo
  {pages} {L201406} (\bibinfo {year} {2026})}\BibitemShut {NoStop}%
\bibitem [{\citenamefont {Palik}\ \emph {et~al.}(1976)\citenamefont {Palik},
  \citenamefont {Kaplan}, \citenamefont {Gammon}, \citenamefont {Kaplan},
  \citenamefont {Wallis},\ and\ \citenamefont {Quinn}}]{InSb_76}%
  \BibitemOpen
  \bibfield  {author} {\bibinfo {author} {\bibfnamefont {E.~D.}\ \bibnamefont
  {Palik}}, \bibinfo {author} {\bibfnamefont {R.}~\bibnamefont {Kaplan}},
  \bibinfo {author} {\bibfnamefont {R.~W.}\ \bibnamefont {Gammon}}, \bibinfo
  {author} {\bibfnamefont {H.}~\bibnamefont {Kaplan}}, \bibinfo {author}
  {\bibfnamefont {R.~F.}\ \bibnamefont {Wallis}},\ and\ \bibinfo {author}
  {\bibfnamefont {J.~J.}\ \bibnamefont {Quinn}},\ }\bibfield  {title} {\bibinfo
  {title} {Coupled surface magnetoplasmon-optic-phonon polariton modes on
  {I}n{S}b},\ }\href {https://doi.org/10.1103/PhysRevB.13.2497} {\bibfield
  {journal} {\bibinfo  {journal} {Phys. Rev. B}\ }\textbf {\bibinfo {volume}
  {13}},\ \bibinfo {pages} {2497} (\bibinfo {year} {1976})}\BibitemShut
  {NoStop}%
\bibitem [{\citenamefont {Ott}\ \emph {et~al.}(2019)\citenamefont {Ott},
  \citenamefont {Messina}, \citenamefont {Ben-Abdallah},\ and\ \citenamefont
  {Biehs}}]{Ott19}%
  \BibitemOpen
  \bibfield  {author} {\bibinfo {author} {\bibfnamefont {A.}~\bibnamefont
  {Ott}}, \bibinfo {author} {\bibfnamefont {R.}~\bibnamefont {Messina}},
  \bibinfo {author} {\bibfnamefont {P.}~\bibnamefont {Ben-Abdallah}},\ and\
  \bibinfo {author} {\bibfnamefont {S.-A.}\ \bibnamefont {Biehs}},\ }\bibfield
  {title} {\bibinfo {title} {Radiative thermal diode driven by nonreciprocal
  surface waves},\ }\href {https://doi.org/10.1063/1.5093626} {\bibfield
  {journal} {\bibinfo  {journal} {Appl. Phys. Lett.}\ }\textbf {\bibinfo
  {volume} {114}},\ \bibinfo {pages} {163105} (\bibinfo {year}
  {2019})}\BibitemShut {NoStop}%
\end{thebibliography}%

\onecolumngrid 
\vspace{0.5cm}
\begin{center}
  \textbf{End Matter} 
\end{center}
\vspace{0.3cm}
\twocolumngrid

\setcounter{equation}{0}
\renewcommand{\theequation}{A\arabic{equation}}

{\it Dielectric properties of the magneto-optic emitter.}
An out-of-plane static magnetic field $\bm{B}=B\hat{z}$ induces a
gyrotropic response in the $xy$-plane of the $n$-doped InSb thermal emitter.
The dielectric tensor takes the form
\begin{equation} \label{MO_z}
  \epsilon_{\rm MO}(\omega) =
  \begin{bmatrix}
    \epsilon_d & -i\epsilon_a & 0 \\
    i\epsilon_a & \epsilon_d & 0 \\
    0 & 0 & \epsilon_p
  \end{bmatrix} .
\end{equation}
Within the Drude-Lorentz framework, these components are
\begin{equation*}
  \epsilon_d = \epsilon_{\infty} \left\{ 1+ \frac{\omega_L^2 -\omega_T^2}{\omega_T^2
  -\omega^2 -i\Gamma\omega} + \frac{\omega_p^2(\omega+i\gamma)}{\omega\big[
  \omega_c^2-(\omega+i\gamma)^2 \big]} \right\}, 
\end{equation*}
\begin{equation*}
  \epsilon_p = \epsilon_{\infty} \left[ 1+ \frac{\omega_L^2 -\omega_T^2}{\omega_T^2
  -\omega^2 -i\Gamma\omega} - \frac{\omega_p^2}{\omega(\omega+i\gamma)} \right], 
\end{equation*}
and
\begin{equation*}
  \epsilon_a = \frac{\epsilon_{\infty}\omega_p^2\omega_c}{\omega\big[ (\omega+i\gamma)^2
  -\omega_c^2 \big]} .
\end{equation*}
For the numerical calculations, the material parameters of $n$-doped InSb are
adopted~\cite{InSb_76, Ott19}: 
high-frequency dielectric constant $\epsilon_\infty = 15.7$,
longitudinal optical phonon frequency $\omega_L = 3.62\times 10^{13}\,{\rm rad/s}$,
transverse optical phonon frequency $\omega_T = 3.39\times 10^{13}\,{\rm rad/s}$, 
phonon damping constant $\Gamma = 5.65\times 10^{11}\,{\rm rad/s}$, 
free-carrier damping constant $\gamma = 3.39\times 10^{12}\,{\rm rad/s}$, and 
plasma frequency $\omega_p = 3.14\times 10^{13}\,{\rm rad/s}$. 
For an applied magnetic field of $B=1\,$T, the cyclotron frequency is $\omega_c =
8.02\times 10^{12}\,{\rm rad/s}$ ($\hbar\omega_c=5.28\,$meV). In the absence of a magnetic
field, the off-diagonal component vanishes, reducing the dielectric tensor to an isotropic
scalar matrix. 
Detailed calculations of the Fresnel reflection coefficients are provided in
Ref.~\cite{GT_WSM}.

{\it Optical response of the monolayer TMDC.} 
The vacuum-TMDC Fresnel reflection and transmission coefficients follow
from the sheet's polarization tensor. For isotropic bands, rotational symmetry allows
$\bm{q}=q\hat{x}$ without loss of generality.
Broken time-reversal symmetry permits an anomalous Hall response,
$\Pi_{xy}^r=-\Pi_{yx}^r$, which mixes $s$ and $p$ polarizations.
Within the random phase approximation (RPA), the retarded
current-current response is given by~\cite{JSW23}
\begin{align} 
  \Pi^{\mathrm{RPA}, r}_{\mu\nu}(\bm{q}, \omega) 
  =& e_0^2 \sum_{n,m,s} \int\frac{d^2\bm{k}}{(2\pi)^2} v_{mn,\mu}^*(\bm{k},\bm{q})
  v_{mn,\nu}(\bm{k},\bm{q}) \notag \\
  & \frac{f(\varepsilon_{n\bm{k},s} +i\eta) - f(\varepsilon_{m,\bm{k} + \bm{q},s}-i\eta)}
  {\hbar\omega + 2i\eta + \varepsilon_{n\bm{k},s} - \varepsilon_{m,\bm{k} + \bm{q},s}} .
  \label{Pi_Lindhard}
\end{align}
The matrix elements of the symmetrized velocity operator are
\begin{equation} 
  \bm{v}_{mn}(\bm{k},\bm{q}) = \langle u_{m,\bm{k}+\bm{q}} | \bm{v}(\bm{k}, \bm{q}) |
  u_{n\bm{k}} \rangle ,
\end{equation}
where $| u_{n\bm{k}} \rangle$ and $| u_{m,\bm{k}+\bm{q}} \rangle$ denote the eigenspinors
for the initial and final states, respectively. They are given by
\begin{align}
  |u_{+, \bm{k}}\rangle 
  &= \frac{1}{\sqrt{2\Delta_{\bm{k},s}\big(\Delta_{\bm{k},s} + h_z\big)}}
  \begin{bmatrix}
    \Delta_{\bm{k},s} + h_z \\
    \lambda_R (k_y - i k_x)
  \end{bmatrix} , \\
  |u_{-, \bm{k}}\rangle 
  &= \frac{1}{\sqrt{2\Delta_{\bm{k},s}\big(\Delta_{\bm{k},s} + h_z\big)}}
  \begin{bmatrix}
    -\lambda_R (k_y + i k_x) \\
    \Delta_{\bm{k},s} + h_z
  \end{bmatrix} ,
\end{align}
with the effective field $h_z=s\lambda_c + M_z$.
The Cartesian components of the symmetrized velocity vertex are given by~\cite{SM}
\begin{align} 
  v_x(\bm{k}, \bm{q}) &= \bar{V}_x(\bm{k}, \bm{q}) \sigma_0 - \frac{\lambda_R}{\hbar}
  \sigma_y , \\
  v_y(\bm{k}, \bm{q}) &= \bar{V}_y(\bm{k}, \bm{q}) \sigma_0 + \frac{\lambda_R}{\hbar}
  \sigma_x ,
\end{align}
where $\bar{\bm{V}}(\bm{k}, \bm{q}) = \hbar ( \bm{k} + \bm{q}/2 ) /  m^*$ is the mean
kinematic velocity. To ensure current conservation by satisfying the gauge-invariance
condition $\Pi^r(\bm{0}, 0) = 0$, the physical polarization function is obtained by
subtracting the diamagnetic contribution from the RPA evaluation, yielding
\begin{equation} 
  \Pi_{\mu\nu}^r(\bm{q}, \omega) = \Pi_{\mu\nu}^{\mathrm{RPA}, r}(\bm{q}, \omega) -
  \Pi_{\mu\nu}^{\mathrm{RPA}, r}(\bm{0}, 0) .
\end{equation}

The retarded photonic Green's function $D_2^r$ at the 2D interface is governed by the
Dyson equation
\begin{equation} \label{Dyson}
  D_2^r = D_0^r + D_0^r \Pi_2^r D_2^r .
\end{equation}
The free-space Green's function expands into the reduced in-plane polarization matrix
$\mathsf{V} = [\hat{s}, \hat{p}]$ as~\cite{GT26-2}
\begin{equation}
  D_0^r = \Upsilon \mathsf{V} \mathsf{V}^T ,
\end{equation}
where the prefactor is $\Upsilon = \mu_0 / (2i\gamma_0)$ with the out-of-plane wavevector
$\gamma_0 = \sqrt{k_0^2 - q^2}$. The reduced polarization vectors are
\begin{equation}
  \hat{s} = 
  \begin{bmatrix} 
    0 \\ -1 
  \end{bmatrix} , \quad
  \hat{p} = 
  \frac{\gamma_0}{k_0} 
  \begin{bmatrix} 
    1 \\ 0 
  \end{bmatrix} ,
\end{equation}
for $\bm{q} =q\hat{x}$. 
Including the boundary scattering, the Green's function $D_2^r$ is expressed via the
transmission matrix $\mathsf{T}_2$ as
\begin{equation}
  D_2^r = \Upsilon \mathsf{V} \mathsf{T}_2 \mathsf{V}^T .
\end{equation} 
Substituting the Green's functions into Eq.~\eqref{Dyson} yields
$\mathsf{T}_2 = (\mathsf{I} - \mathsf{M})^{-1}$ with $\mathsf{I}$ the identity matrix and 
\begin{equation}
  \mathsf{M} = \Upsilon \mathsf{V}^T \Pi_2^r \mathsf{V} 
  \equiv
  \begin{bmatrix} 
    \alpha_s & \alpha_h \\ -\alpha_h & \alpha_p 
  \end{bmatrix} ,
\end{equation}
with the components given by
\begin{equation}
  \alpha_s = \frac{-i \mu_0 \Pi_{yy}^r}{2 \gamma_0} , \quad
  \alpha_p = \frac{-i \gamma_0 \Pi_{xx}^r}{2 \epsilon_0 \omega^2} , \quad
  \alpha_h = \frac{-i \mu_0 \Pi_{xy}^r}{2 k_0} .
\end{equation}
The transmission matrix is thus expressed as
\begin{equation}
  \mathsf{T}_2 \equiv 
  \begin{bmatrix}
    t^{s}_2 & t^{sp}_2 \\
    t^{ps}_2 & t^{p}_2
  \end{bmatrix}
  = \frac{1}{\Delta} 
  \begin{bmatrix} 
    1 - \alpha_p & \alpha_h \\ 
    -\alpha_h & 1 - \alpha_s 
  \end{bmatrix} ,
\end{equation}
with $\Delta = (1 - \alpha_s)(1 - \alpha_p) + \alpha_h^2$.
The reflection matrix is then obtained as
\begin{equation}
  \mathsf{R}_2 \equiv 
  \begin{bmatrix}
    r^s_2 & r^{sp}_2 \\
    r^{ps}_2 & r^p_2
  \end{bmatrix}
  =   
  \begin{bmatrix}
    1 & 0 \\
    0 & -1
  \end{bmatrix} 
  (\mathsf{T}_2-\mathsf{I} ) ,
\end{equation}
with its elements as
\begin{align}
  & r^s_2 = \big[\alpha_s(1 - \alpha_p) - \alpha_h^2 \big] / \Delta , \\
  & r^p_2 = \big[-\alpha_p(1 - \alpha_s) + \alpha_h^2 \big] / \Delta , \\
  & r^{sp}_2 = r^{ps}_2 = \alpha_h / \Delta .
\end{align}

Under time-reversal symmetry ($\alpha_h=0$), cross-polarization
coefficients vanish, and the transverse and longitudinal responses
determine the $s$- and $p$-reflection coefficients, respectively:
\begin{equation}
  r^s_2=-\frac{i\mu_0\Pi_{yy}^r}{2\gamma_0+i\mu_0\Pi_{yy}^r}, \qquad
  r^p_2=\frac{i\gamma_0\Pi_{xx}^r}
  {2\epsilon_0\omega^2+i\gamma_0\Pi_{xx}^r},
\end{equation}
where $\Pi_{\mu\mu}^r\equiv\Pi_{\mu\mu}^r(q\hat{x},\omega)$.
The transmission coefficients are $t^s_2=1+r^s_2$ and $t^p_2=1-r^p_2$.

{\it Near-field heat flux and spin yield.}
For the nonchiral configuration, retaining only the $p$-polarized evanescent contribution
gives the heat flux absorbed by the TMDC~\cite{review07,review15,review18,review21}:
\begin{equation} \label{heat}
  J = \int_0^\infty\frac{d\omega}{2\pi} \int_{q>k_0}\frac{d^2\bm{q}}{(2\pi)^2}
  \hbar\omega\,\mathcal{T}(\bm{q},\omega)N_d(\omega),
\end{equation}
where the photonic transmission coefficient is 
\begin{equation} \label{trans}
  \mathcal{T}(\bm{q},\omega)
  = 4\,{\rm Im}(r_1^p){\rm Im}(r_2^p)e^{-2|\gamma_0|d} / |\Delta_p|^2 .
\end{equation}
The net photon absorption rate per unit area, $\dot N_{\rm ph}$,
is obtained by replacing $\hbar\omega$ in Eq.~\eqref{heat} with unity.
For $\eta=6\,$meV, we obtain
$J \approx 1.05\times 10^4\,\mathrm{W / m^2}$ and
$\dot{N}_{\rm ph}\approx 3.3\times10^{24}\,\mathrm{m^{-2}\,s^{-1}}$,
both nearly independent of $M_z$ over the range shown in Fig.~\ref{fig4}.
Their ratio gives the mean absorbed photon energy: $\langle \hbar\omega \rangle = J /
\dot{N}_{\rm ph} \approx 20\,$meV,
consistent with the energy scale of the InSb resonances near
$17$ and $26\,$meV.
The spin yield is defined as 
\begin{equation} \label{Ys}
  Y_s = |I_z^{\rm nc}| / (\hbar\dot{N}_{\rm ph}) ,
\end{equation}
where $I_z^{\rm nc}$ is given by Eq.~\eqref{Iz_nc}.
This dimensionless quantity measures the magnitude of the net
out-of-plane spin angular momentum transferred per net absorbed
photon, in units of $\hbar$. As shown in Fig.~\ref{fig4}(b), 
at $M_z=1\,$meV and $\eta=6\,$meV, $Y_s\approx 4.4 \times 10^{-5}$.
This small yield reflects the weak spin transfer relative
to the overall photon absorption.

\end{document}